\documentclass[a4paper,fleqn,usenatbib]{mnras}

\usepackage{newtxtext,newtxmath}

\usepackage[T1]{fontenc}
\usepackage{ae,aecompl}

\usepackage{graphicx}	
\usepackage{amsmath}	

\usepackage{enumitem}
\setlist[itemize]{leftmargin=*}

\usepackage{epsfig}
\usepackage{multirow}
\usepackage{float}
\usepackage{footmisc}

\def\apjl{ApJL}
\def\apjs{ApJS}
\def\aap{A\&A}

\def\xmm{{\it XMM-Newton}}

\def\nustar{{\it NuSTAR}}
\def\swift{{\it Swift}}

\def\rx04{RX04}
\def\src1{RX01}
\def\rej1034{RE10}
\def\1h07{1H07}
\def\pg12{PG1244+026}
\def\grs1915{GRS 1915+105}
\def\ph1092{PH10}
\def\phl1811{PH18}

\def\rxj0134{RX J0134.2-4258}

\title[RX J0134.2-4258 -- II. a WLS Linking to the WLQ]{The Extreme Super-Eddington NLS1 RX J0134.2-4258 -- II. A Weak-Line Seyfert Linking to the Weak-Line Quasar}

\author[C. Jin, et al.]{Chichuan Jin$^{1,2}$\thanks{E-mail: ccjin@nao.cas.cn},
Chris Done$^{3}$,
Martin Ward$^{3}$,
Francesca Panessa$^{4}$,
Bo Liu$^{1}$,
He-Yang Liu$^{1}$
\smallskip
\\
$^{1}$National Astronomical Observatories, Chinese Academy of Sciences, 20A Datun Road, Beijing 100101, China\\
$^{2}$School of Astronomy and Space Sciences, University of Chinese Academy of Sciences, 19A Yuquan Road, Beijing 100049, China\\
$^{3}$Centre for Extragalactic Astronomy, Department of Physics, University of Durham, South Road, Durham DH1 3LE, UK\\
$^{4}$INAF - Istituto di Astrofisica e Planetologia Spaziali (IAPS-INAF), Via del Fosso del Cavaliere 100, I-00133 Roma, Italy\\
}

\date{prepared for MNRAS}

\pubyear{2022}

\begin{document}
\label{firstpage}
\pagerange{\pageref{firstpage}--\pageref{lastpage}}
\maketitle

\begin{abstract}
\rxj0134\ is one of the most super-Eddington narrow-line Seyfert 1 (NLS1) galaxies, on which we conducted a monitoring campaign from radio to X-rays. In this paper, we present a detailed analysis of its optical/UV spectra and broadband spectral energy distribution (SED). Our study shows that the preferred black hole mass of \rxj0134\ is $M_{\rm BH} \sim 2 \times 10^{7}~M_{\odot}$, giving a mass accretion rate through the outer disc of $\dot{m}_{\rm out} \sim 20$ (assuming zero spin), compared to the observed luminosity ratio $L_{\rm bol}/L_{\rm Edd} \sim 6$. This reduction in radiative efficiency is expected for super-Eddington flows, as power can be lost via advection and/or disc winds. We find that the optical/UV lines of \rxj0134\ resemble those from weak-like quasars (WLQs), as it has notably weak C {\sc iv} and N {\sc v} emission lines. It also has drastic X-ray variability, again similar to that recently observed in some other WLQs. However, WLQs have systematically higher masses ($\gtrsim 10^8~M_{\odot}$), and lower Eddington ratios ($\dot{m}_{\rm out} \sim 1$) than \rxj0134. We compare instead to the most extreme NLS1s, with similarly large $\dot{m}_{\rm out}$ but smaller masses. These show similarly large reductions in radiative efficiency but their UV lines are not similarly wind-dominated. We suggest a new category of weak-line Seyfert (WLS) galaxies to describe sources like \rxj0134, and interpret its (so far unique) properties in a model, where the lower-disc-temperature in the higher-mass black holes leads to the UV-line-driving mechanism, which enhances the super-Eddington radiation-pressure-driven wind.
\end{abstract}




\section{Introduction}
\label{sec-intro}
\subsection{Narrow-line Seyfert 1 Galaxies}
Active galactic nuclei (AGN) are powered by accretion onto a super-massive black hole (SMBH). This converts some fraction of the gravitational potential energy into radiation, powering the observed activity. The multi-wavelength properties of AGN are mainly determined by three parameters, namely the black hole mass, black hole spin and mass accretion rate. The inclination angle also plays a significant role in observations (e.g. \citealt{Luo.2015, Jin.2017b}). Narrow-line Seyfert 1 (NLS1) galaxies are a subtype of AGN characterized by relatively narrow broad lines such as H$\beta$ and relatively weak narrow lines such as [O {\sc iii}]$\lambda$5007 (\citealt{Osterbrock.1985}; \citealt{Boroson.2002}). Comparing with the entire AGN population, NLS1s tend to have small black hole masses of $10^{6-7}~M_{\odot}$ and high mass accretion rates  (e.g. \citealt{Pounds.1995}; \citealt{Mathur.2001}; \citealt{Boroson.2002}; \citealt{Jin.2012a}).

In the X-ray band, it is common to observe a strong soft X-ray excess in NLS1s (e.g. \citealt{Boller.1996, Brandt.1997}), which can often be modelled with an ionized disc reflection component (e.g. \citealt{Miniutti.2004, Ross.2005, Crummy.2006, Fabian.2013}), and/or a separate warm Comptonisation component (e.g. \citealt{Laor.1997}; \citealt{Magdziarz.1998, Done.2012, Jin.2013, Jin.2016, Jin.2017a, Jin.2021}). Complex absorption (partially ionised material partially covering the source) can also shape the soft X-ray emission in some AGN (e.g. \citealt{Miller.2007, Turner.2007, Tatum.2012}).

NLS1s themselves form two subtypes, including the X-ray {\it simple} NLS1s and X-ray {\it complex} NLS1s (\citealt{Gallo.2006}). The X-ray {\it simple} NLS1s have smooth and steep X-ray spectra, while the X-ray {\it complex} NLS1s show more complicated absorption and emission features. Meanwhile, NLS1s with high mass accretion rates, especially super-Eddington, are likely to have a geometrically-thick (i.e. puffed-up) inner disc structure and disc wind (\citealt{Ohsuga.2011, Takeuchi.2014, Jiang.2016}), which can obscure the intrinsic X-ray emission and introduce additional spectral complexities and variability (e.g. \citealt{Hagino.2016, Done.2016, Jin.2017b, Parker.2021}). Therefore, the difference between X-ray {\it simple} and {\it complex} NLS1s can be explained by their different inclination angles, which lead to different line-of-sight to the X-ray corona (\citealt{Done.2016, Jin.2017b}). Supporting evidence for these subtypes being intrinsically the same is that their optical/UV emission is the same,
suggesting that their intrinsic disc properties should indeed be similar (\citealt{Done.2016}).

\subsection{Weak-line Quasars}
A similar physical scenario of a puffed-up inner disc with significant winds is proposed to explain the properties of weak-line quasars (WLQs, e.g. \citealt{Fan.1999, Plotkin.2010, Wu.2011, Wu.2012, Luo.2015, Ni.2018}). WLQs are characterized by their weak UV high-ionization emission lines, e.g. rest-frame equivalent width (REW) of C {\sc iv} 
$\lesssim 10$ \AA, and/or REW of Ly $\alpha~+$ N {\sc v} 
is $\lesssim 15$ \AA\ (e.g. \citealt{Ni.2018}). Winds are clearly indicated as the peak of the weak C {\sc iv} lines is often highly blue-shifted (\citealt{Richards.2011, Rankine.2020}).
WLQs have high black hole masses of $10^{8-9}M_\odot$, and also fairly high but not extreme mass accretion rates of $L_{\rm bol}/L_{\rm Edd}\sim 1$ (e.g. \citealt{Luo.2015}). The empirical $\alpha_{\rm ox}-L_{\rm 2500\text{\AA}}$ relation (e.g. \citealt{Lusso.2016}) implies that these are somewhat X-ray weak compared to less luminous quasars, but $\sim$35\% of the WLQ population show X-ray emission which is at least a factor 6 below this expectation (\citealt{Pu.2020}). 
This fraction of X-ray weakness is significantly higher than in the non-WLQ AGN population. A plausible explanation is that the high Eddington ratio causes the inner accretion disc of a WLQ to puff up, which partially shields the X-ray emission from near the black hole. Then the observed X-ray emission will depend on the viewing angle, in which case an X-ray weak WLQ will have a higher inclination angle, so that the line-of-sight to the X-ray corona is obscured by the geometrically thick inner disc (e.g. \citealt{Wu.2011, Luo.2015, Ni.2018}).
 
Therefore, WLQs and super-Eddington NLS1s share some similar 
properties (\citealt{Leighly.2007b, Jin.2017b}), yet they do also differ significantly in black hole masses and, more importantly, in Eddington ratios, so the disc structure and geometry need not be the same. A more detailed comparison between these two AGN populations would allow us to better understand the evolution of super-Eddington accretion flows with black hole mass and mass accretion rate.

\subsection{The Multi-wavelength Campaign on \rxj0134}
We conduct a new multi-wavelength campaign from radio to hard X-rays to observe one of the most extreme super-Eddington NLS1s, namely \rxj0134 in order to deepen our understanding about super-Eddington accretion. 
This campaign involves new observations with \xmm, \nustar, \swift, {\it ATCA} and the 2.3-m telescope in the Sliding Spring Observatory (SSO), as well as a large set of archival multi-wavelength data (see Section~\ref{sec-obs} and \citealt{Jin.2022}, here after: Paper-I).

\rxj0134 was discovered by \citet{Voges.1999} in the {\it ROSAT} all sky survey. Its key properties are summarized below, while a more detailed introduction can be found in Paper-I. This NLS1 lies at the redshift of 0.237, and it appears as an unresolved source in optical. It has a black hole mass of $M_{\rm BH}\simeq1.5\times10^{7}M_{\odot}$ and an extremely high Eddington ratio of $L_{\rm bol}/L_{\rm Edd}\simeq10.0$ (\citealt{Grupe.2010}). 
It has a steep hard X-ray slope ($\Gamma \simeq 2.2$, Paper-I), typical of NLS1, but 
has only an extremely weak soft X-ray excess, which is both peculiar and puzzling. In addition, it also exhibits drastic X-ray variability in terms of both spectral shape and flux (Paper-I). Its optical/UV properties such as the extremely weak [O {\sc iii}] $\lambda$5007 and blue-shifted C {\sc iv} were shown to be similar to the WLQ PHL 1811 by \citet{Leighly.2007b}.

The latest simultaneous \xmm\ and \nustar\ observations in our campaign caught \rxj0134\ in its one of the lowest X-ray flux states in history, thus we conducted a detailed X-ray spectral-timing analysis. As shown in Paper-I, we found that the time-average X-ray spectra in the low-flux state has excess flux above 4 keV, which is lagged by $\sim$4 ks behind the soft X-rays. The spectral-timing properties in both low and high-flux states can be well modelled under the warm Comptonisation plus a distant neutral reflection scenario, or by a partial covering absorption scenario. Both scenarios require a clumpy disc wind in this super-Eddington accretion system.

Here we perform a detailed 
multi-wavelength study from infrared, through optical/UV and then to X-rays to 
provide independent constraints on the global properties of the accretion flow. For example, the optical/UV continuum can be used to measure the mass accretion rate through the outer disc (e.g. \citealt{Davis.2011, Done.2016}). The optical/UV emission/absorption lines can provide information about the broad-line region and outflows (e.g. \citealt{Bottorff.1997, Pancoast.2011, Pancoast.2014, Grier.2017, Li.2018}), and can also be used to measure virial black hole mass (e.g. \citealt{Peterson.2004, Vestergaard.2006, Peterson.2014, Du.2019}). The infrared emission can be used to constrain the properties of the dusty torus (e.g. \citealt{Fuller.2016, Collinson.2017, Martinez.2017, Landt.2019}). The broadband spectral energy distribution (SED) can be used to estimate the black hole mass and Eddington ratio (e.g. \citealt{Jin.2012a, Jin.2016, Jin.2017b}), which can then be used to measure the global radiative efficiency ($\mu$, e.g. \citealt{Davis.2011}). In this work, we collate a large multi-wavelength dataset to study \rxj0134.

\subsection{The Scope of This Paper}
This paper presents a detailed study on the optical/UV and broadband SED properties of \rxj0134, as well as a detailed comparison with some representative super-Eddington NLS1s and WLQs. We will suggest a new category of weak-line Seyfert (WLS) galaxies, and demonstrate that \rxj0134\ is an archetypal WLS.

The structure of this paper is as follows. Firstly, we describe the multi-wavelength datasets used in this work, and then briefly describe the data reduction procedures. Then we present a detailed estimate of the black hole mass of \rxj0134\ because it is a key parameter. Section 4 presents a detailed multi-component broadband SED modelling, in order
to derive key parameters such as the bolometric luminosity, mass accretion rate and Eddington ratio. In Section 5, we first compare \rxj0134\ with WLQs, and propose it as an archetypal WLS, i.e. a new category of AGN. Then we use a small sample to conduct a more general comparison between the super-Eddington NLS1 population and the more typically Eddington WLQ population.  In Section 6, we propose a picture for super-Eddington accretion flows with different parameters. We show how the disc properties, such as the disc structure, wind and global radiative efficiency, may depend on the black hole mass and mass accretion rate. Section 7 summarizes the main results of this paper. A detailed optical/UV spectral analysis is presented in the appendix.

We adopt a flat universe model throughout this work, with the Hubble constant H$_{0} = 72$ km s$^{-1}$ Mpc$^{-1}$, $\Omega_{\Lambda} = 0.73$ and $\Omega_{\rm M} = 0.27$.

\begin{table}
\centering
   \caption{The multi-wavelength dataset of \rxj0134\ used in this work. $T_{\rm obs}$ is the total observing time. For \nustar\ the Earth occultations and south atlantic anomaly passages have been excluded. A complete list of all the observations used by this research project can be found in Paper-I.}
     \begin{tabular}{@{}lccr@{}}
     \hline
    Instrument & Obs-Date & $T_{\rm obs}$ & Waveband \\
     & & (ks) &\\
    \hline
    \multicolumn{4}{c}{New Observations} \\
    {\it NuSTAR} FPMA/FPMB& 2019-12-19 & 98.3 & Hard X-ray \\
    \xmm\ EPIC/OM & 2019-12-19 & 134.3 & X-ray/UV \\
    {\it Swift} XRT/UVOT & 2019-12-19 & 1.6 & X-ray/UV/Optical \\
    {\it SSO 2.3-m Telescope} & 2019-12-19 & 1.8 & Optical \\
    \hline
     \multicolumn{4}{c}{Archival Observations} \\
    \xmm\ EPIC/OM & 2008-12-11 & 32.1 & X-ray/UV/Optical \\
    {\it HST} FOS & 1996-09-21 & 1.7 & UV (G130H) \\
    {\it HST} FOS & 1996-09-21 & 2.1 & UV (G130H) \\
    {\it HST} FOS & 1996-09-21 & 0.2 & UV (G160L) \\
    {\it HST} FOS & 1996-09-21 & 1.5 & UV (G190H) \\
    {\it HST} FOS & 1996-09-21 & 1.2 & UV (G270H) \\
    {\it HST} FOS & 1996-09-21 & 1.0 & Optical (G400H) \\
    {\it HST} FOS & 1996-09-21 & 0.6 & Optical (G570H) \\
    {\it WISE} & 2010-06-20 & -- & Infrared (Band 1-4) \\
    {\it 2MASS} & 1999-08-27 & -- & Infrared ({\it J}, {\it H}, {\it K})\\
    \hline
     \end{tabular}
\label{tab-obs}
\end{table}

\section{Observations and Data Reduction}
\label{sec-obs}
We use a large number of observations, from both our new campaign and previous observations. These datasets are listed in Table~\ref{tab-obs}. The \xmm\ and \nustar\ data have been used in Paper-I for detailed X-ray spectral-timing analysis, where their data reduction are described in more detail.

\subsection{X-ray Observations}
There are two \xmm\ (\citealt{Jansen.2001}) observations for \rxj0134, whose observation dates differ by 11 years. The first observation in 2008 is referred to as Obs-1, and the second in 2019 is Obs-2. These observations also have 
simultaneous optical/UV data from various filters with the optical monitor (OM). We summarize the data reduction procedures below. The data are downloaded from the \xmm\ Science Archive (XSA), and reprocessed with the {\tt epproc} and {\tt empproc} tasks in the \xmm\ Science Analysis System (SAS v18.0.0). The source extraction region was chosen to be a circle of 35 arcsec radius, and no pile-up effect was detected during the two observations. The source and background spectra were extracted with the {\tt evselect} task, and the response and auxiliary files were produced by the {\tt rmfgen} and {\tt arfgen} tasks. The data obtained by the optical monitor (OM) were reprocessed with the {\tt omichain} task.

The \nustar\ (\citealt{Harrison.2013}) observation of \rxj0134\ was conducted simultaneously with the \xmm\ observation in 2019. The {\tt nupipline} task inside the HEASoft package (v6.27.2, \citealt{Blackburn.1995}) was used to reprocess the data. The source extraction region was chosen to be a circle with 1 arcmin radius, and the background was extracted from a nearby circular source-free region with the same radius. The {\tt nuproducts} task was used to extract the spectra.

There are 51 \swift\ (\citealt{Gehrels.2004}) observations on \rxj0134\ from 2019-12-19 to 2021-08-22. In this work we only use the observation conducted on 2019-12-31, because it is simultaneous with the \xmm\ and \nustar\ observations. A complete analysis of all the \swift\ observations will be present in a following paper (Panessa et al. in preparation, hereafter: Paper-III). Six filters were used in the \swift\ Ultra-violet Optical Telescope (UVOT) during this \swift\ observation (i.e. UVW2, UVM2, UVW1, U, B and V). The HEASsoft (v6.27.2) package was used to reduce the data. The \swift\ X-ray Telescope (XRT) data were reprocessed with the {\tt xrtpipeline}. The source spectrum was extracted from a circular region of 30 arcsec radius. For the UVOT photometric data, a circular aperture of 5 arcsec radius was adopted. Background was chosen from nearby source-free regions with larger areas. We also ran the standard sensitivity check for UVOT, in order to ensure that the data are not affected by the regions on the detector where the throughputs are degraded due to the contamination of dust/debris (\citealt{Edelson.2015}).

\subsection{Optical/UV/Infrared Observations}
\label{sec-optuv-spec}
{\it Hubble} Space Telescope ({\it HST}) observed \rxj0134\ in 1996 with the Faint Object Spectrograph (FOS), which covered the spectral range of 970 -- 5500 \AA\ in the AGN rest-frame. The calibrated data were downloaded from the Mikulski Archive for Space Telescopes (MAST), from which the spectra were extracted with the IRAF/STSDAS tasks following the standard procedure\footnote{https://www.stsci.edu/instruments/wfpc2/Wfpc2\_dhb/intro\_ch36.html}. 
We obtained a new optical spectrum of \rxj0134\ with the SSO 2.3-m telescope on 2019-12-19. Infrared photometry from {\it WISE} (band: 1 -- 4) and {\it 2MASS} (band: {\it J}, {\it H}, {\it K}) were downloaded from the NASA/IPAC Infrared Science Archive (IRSA).

\begin{figure*}
\centering
\begin{tabular}{c}
\includegraphics[trim=0.2in 0.3in 0.0in 0.0in, clip=1, scale=0.58]{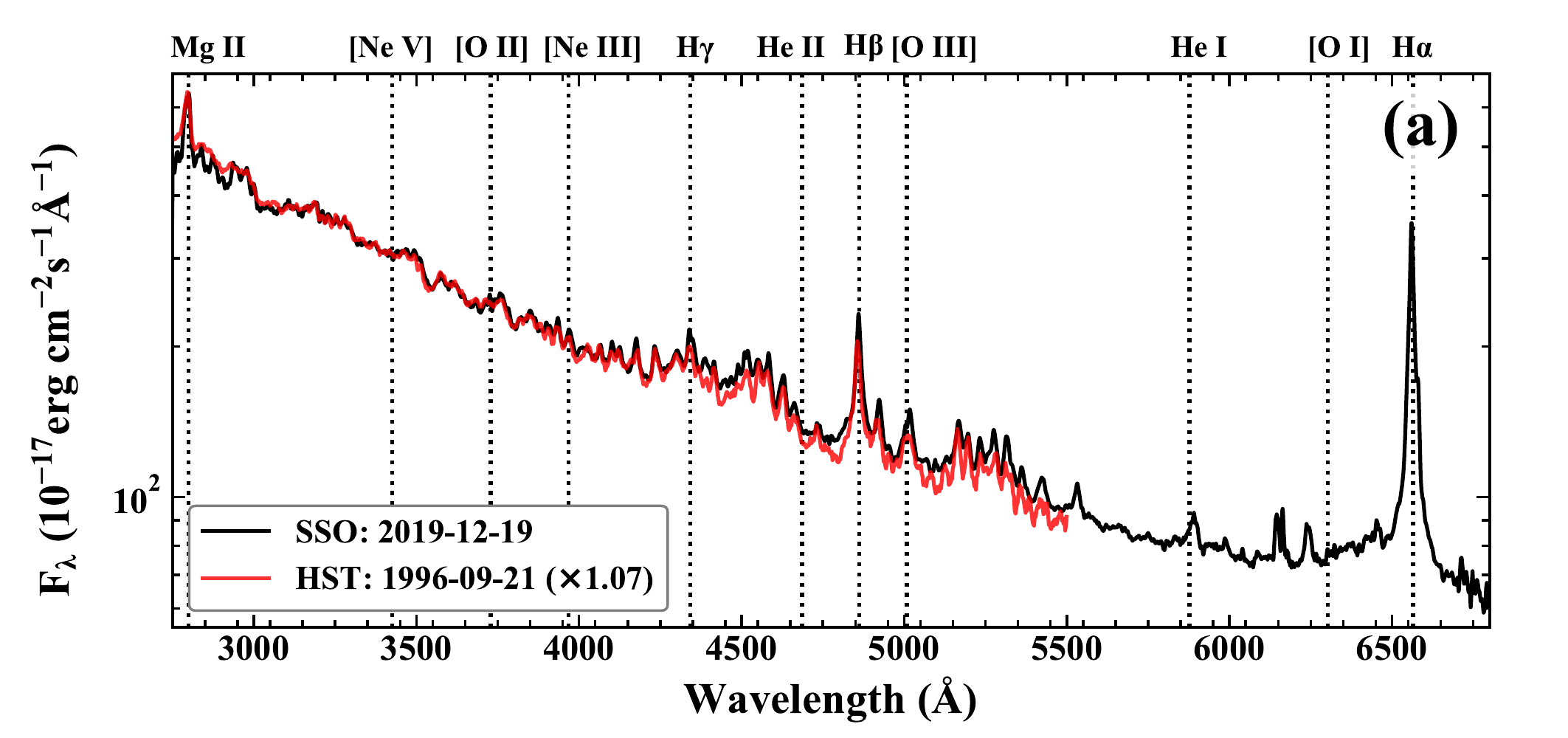} \\
\includegraphics[trim=0.2in 0.0in 0.0in -0.2in, clip=1, scale=0.58]{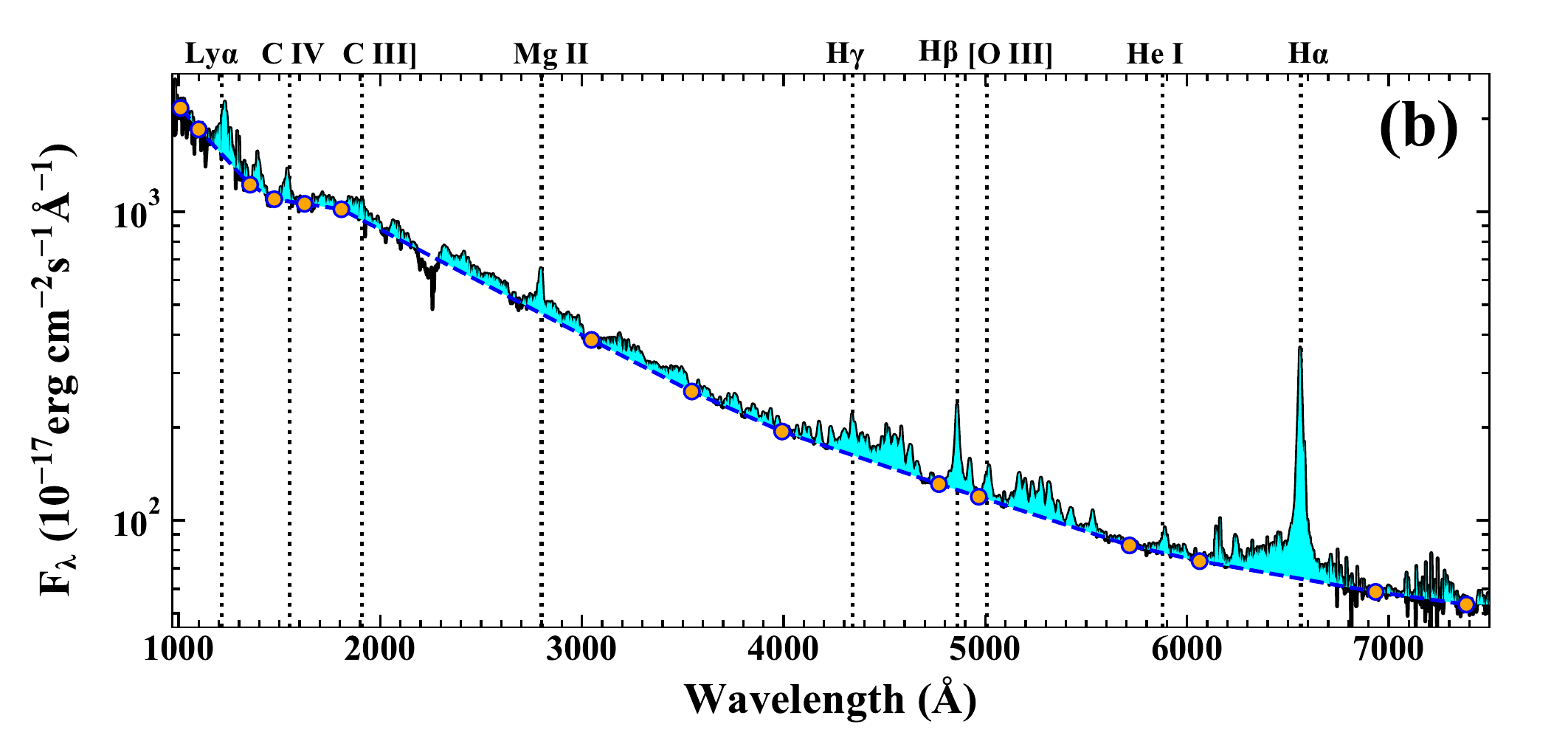} \\
\end{tabular}
\caption{Panel-a: comparison between the {\it HST}/FOS spectrum in 1996 (red) and the SSO spectrum in 2019 (black). The {\it HST} spectrum is scaled up by a factor of 1.07 to obtain the best match to the SSO spectrum. Panel-b: the combined optical/UV spectrum of \rxj0134, based on the {\it HST} spectrum ($\times1.07$) below 4000\AA\ and the SSO spectrum above 4000\AA. The spectra have been de-redshifted to the AGN rest frame. The orange circles indicate the positions used to define the underlying continuum (blue dash line). The cyan region indicates emission lines superposed on the continuum. All the spectra have been de-reddened for the Galactic reddening before redshift correction.}
\label{fig-optspec1}
\end{figure*}

Then we analyze the SSO optical and {\it HST} UV spectra, 
fitting for the lines and continuum components. Details are given in Appendix~\ref{sec-optuvfit}, with spectra shown in Figure~\ref{fig-optspec1}a,
de-reddened with $E(B-V)=0.0144$ for the Fitzpatrick \& Massa (2007) reddening curve for $R_{\rm V}$ = 3.1, and de-redshifted for $z=0.237$.
These data are separated by 23 years but 
the overall difference of normalization is only 7\%. Below 4000\AA, the two spectra match almost perfectly after removing this 7\% difference, while above 4000\AA\ the {\it HST} spectrum is weaker by another 5\%. Therefore, we can connect the {\it HST} spectrum (scaled up by 1.07) and the SSO spectrum at 4000\AA\ to derive a broad optical/UV spectrum, which is shown in Figure~\ref{fig-optspec1}b.

The optical spectrum of \rxj0134\ resembles a typical NLS1 galaxy. According to the empirical eigenvector 1 of AGN (e.g. \citealt{Boroson.2002, Jin.2012c}), the strong Fe {\sc ii} and weak [O {\sc iii}]$\lambda$5007 lines imply that \rxj0134\ should have a very high mass accretion rate. Another key property of \rxj0134\ is that its UV spectrum has very weak and blue-shifted C {\sc iv} and Ly$\alpha$+N{\sc v} emission lines, like WLQs. We performed a detailed multiple Gaussian+Lorentzian profile decomposition for different emission lines. The methods and results are described in Appendix~\ref{sec-optuvfit}. The optical/UV line decomposition and best-fit parameters can be found in Figures~\ref{tab-optfit}, \ref{tab-uvfit} and Tables~\ref{tab-optfit}, \ref{tab-uvfit}.

\section{The Black Hole Mass}
\label{sec-bhmass}
\citet{Grupe.2010} reported a virial mass of $M_{\rm BH}~=~1.47\times10^{7}M_{\odot}$ for \rxj0134, which is based on the single-epoch H$\beta$ full width at half-maximum (FWHM) of 1160 km s$^{-1}$, and the radius-luminosity (R-L) relation reported by \citet{Kaspi.2000}. We use the latest SSO optical spectrum and measure the H$\beta$ FWHM to be 1140 $\pm$ 20 km s$^{-1}$ for the Lorentzian decomposition, and 1410 $\pm$ 70 km s$^{-1}$ for the Gaussian decomposition. The monochromatic luminosity at the rest-frame 5100 \AA\ is measured to be $(8.86\pm0.92)\times10^{44}$ erg s$^{-1}$. For the two H$\beta$ FWHM measurements and using a later R-L relation reported by \citet{Vestergaard.2006}, we obtain a black hole mass of $M_{\rm BH}~=~(3.1-4.8)\times10^{7}M_{\odot}$.

However, recent reverberation mapping studies have shown that for super-Eddington AGN, the observed radius of the broad line region (BLR) is smaller than expected from the classic R-L relation (\citealt{Du.2018}). This is likely due to changes in the disc structure and radiation as the flow becomes super-Eddington. Firstly the accretion flow has intrinsically lower radiative efficiency than a standard disc due to advection and/or winds, and secondly the inner disc may become geometrically thick, which, together with any wind, can provide a shielding mechanism for the ionization of BLR (e.g. \citealt{Abramowicz.1988, Wang.2003, Jiang.2014, Done.2016, Jin.2016, Jin.2017b}). As a result, previous versions of R-L relation can lead to an over-estimate of the black hole mass in the super-Eddington regime. 

Recently, \citet{Du.2019} reported a new R-L relation which includes the Fe {\sc ii} to H$\beta$ (broad component) flux ratio ($R_{\rm FeII}$) as an additional parameter. This new relation provides lower mass estimates than traditional relations for super-Eddington AGN. For \rxj0134, $R_{\rm FeII}$ is found to be 1.74 $\pm$ 0.15 (see Section~\ref{sec-opt-spec}), which is larger than most of the sources in \citet{Du.2019}.
Then it is necessary to choose a value for the 
virial factor $f_{\rm BLR}$, which depends on the morphology of the host galaxy. Since the host galaxy of \rxj0134\ cannot be resolved, different $f_{\rm BLR}$ values need to be tried to understand its impact on the mass estimate (see Table~\ref{tab-mass-mdot}). \citet{Ho.2014} reported $f_{\rm BLR}=0.7\pm0.2$ for AGN in pseudo-bulges, which leads to a mass of  $(0.79\pm0.25)\times10^{7}M_{\odot}$. Then for AGN in classic bulges or ellipticals with $f_{\rm BLR}=1.5\pm0.4$, the mass is $(1.70\pm0.54)\times10^{7}M_{\odot}$. If we choose $f_{\rm BLR}=1.12$ for the sample of 93 NLS1s reported by \citet{Woo.2015}, then the mass is found to be $(1.27\pm0.22)\times10^{7}M_{\odot}$. Therefore, it is clear that the uncertainty of $f_{\rm BLR}$ affects the single-epoch virial mass significantly. The intrinsic scatter of 0.2 dex of this new R-L relation introduces further uncertainty, and so there is no significant difference between these virial masses.

The rapid X-ray variability provides an independent method to estimate the black hole mass.
Various studies have shown that the mass scales with the X-ray rms (\citealt{Lu.2001, Zhou.2010, Ponti.2012, Jin.2016}). We calculate the X-ray rms for different variability timescales, leading to a mass range of $(0.8-2.5)\times10^{7}M_{\odot}$ and a mean value of $1.7\times10^{7}M_{\odot}$ (see Paper-I). This mass estimate is subject to an intrinsic scatter of 0.7 dex.

Overall, a typical mass estimate of $M_{\rm BH}\sim 2\times10^{7}M_{\odot}$ for \rxj0134\ should be statistically consistent with all the mass estimates presented above.

\begin{table}
\centering
   \caption{Different mass estimates for \rxj0134, and the corresponding values for some other key parameters, including the mass accretion rates ($\dot{m}_{\rm out}$), Eddington ratio ($L_{\rm bol}/L_{\rm Edd}$) and radiative efficiency ($\mu$). The best-fit SED for Obs-1 assumes $a_{*}=0$ and $\mu_0=0.057$, and has $L_{\rm bol}=1.63\times10^{46}$ erg s$^{-1}$. Typical uncertainties are provided for $M_{\rm BH}$ and propagated into $\dot{m}_{\rm out}$, but the intrinsic 0.2 dex scatter is not included. Systematic uncertainties should dominate $L_{\rm bol}$ and $\mu$, so their errors are not provided.}
     \begin{tabular}{@{}lccccc@{}}
    \hline
    Method & $M_{\rm BH}$ & $L_{\rm bol}/L_{\rm Edd}$ & $\dot{m}_{\rm out}$ & $\mu/\mu_{\rm 0}$ & $\mu$ \\
    & ($10^{7}M_{\odot}$) & & \\
    \hline
    Best-fit SED & 2.00 fixed & 6.3 & 20.6 $^{+0.3}_{-0.6}$ & 0.31 & 0.017 \\
    X-ray Rms & 1.70 $\pm$ 0.80 & 7.4 & 28.5 $^{+0.4}_{-0.8}$ & 0.26 & 0.015 \\
    \multicolumn{3}{@{}l}{R-L Relation from \citet{Du.2019}} \\
    ($f_{\rm BLR}=0.70$) & 0.79 $\pm$ 0.25  & 15.9 & 132.0 $^{+1.9}_{-3.8}$ & 0.12 & 0.007 \\
    ($f_{\rm BLR}=1.12$) & 1.27 $\pm$ 0.22 & 9.9 & 51.1 $^{+0.7}_{-1.5}$ & 0.19 & 0.011 \\
    ($f_{\rm BLR}=1.50$) & 1.70 $\pm$ 0.54  & 7.4 & 28.5 $^{+0.4}_{-0.8}$ & 0.26 & 0.015 \\
    \hline
     \end{tabular}
\label{tab-mass-mdot}
\end{table}

\section{Multi-wavelength Properties}
\label{sec-multiwavelength}
\subsection{Broadband Spectral Energy Distribution}
\label{sec-sed}
\subsubsection{Preparation of the Multi-wavelength Data}
The study of spectral energy distribution (SED) can provide crucial information about the accretion system, such as the black hole mass and spin, mass accretion rate, the energy budget in different wavebands and spectral components (e.g. \citealt{Jin.2012a, Done.2012, Done.2013}). The abundant multi-wavelength data collected from our new observations and public data archives allow us to reconstruct the broadband SED of \rxj0134.

The dataset used to construct the SED is listed in Table~\ref{tab-obs}, which includes \xmm\ EPIC-pn and five OM filters (UVW2, UVM2, UVW1, U and B) from Obs-1.
We neglect Obs-2, with its factor of $\sim$ 4 lower EPIC-pn count rate, as it has a more complex X-ray shape which is most likely due to absorption variability rather than intrinsic spectral change (see Paper-I). We note that the 
corresponding simultaneous UV fluxes from the OM UVW1 filter are more similar, with Obs-2 being 16 per cent brighter than in Obs-1.
The remaining datasets used are \swift\ XRT and UVOT, {\it ROSAT} PSPCB spectrum, {\it HST} FOS spectra, {\it 2MASS} and {\it WISE} photometric points. 

\rxj0134\ shows all kinds of emission lines in its optical/UV spectrum, which need to be removed so that the continuum can be used for the SED fitting. These lines will also contribute to the optical/UV photometry, thus we need to perform corrections for all the optical/UV photometric fluxes. We visually inspect the optical/UV spectrum, and choose a series of data points to define the underlying continuum, as shown by the yellow points and blue dash line in Figure~\ref{fig-optspec1}b. The cyan region between this continuum and the observed spectrum is considered to come from emission lines. The line-free continuum is converted into {\sc xspec}-readable spectral file for the SED fitting. We also calculate the fraction of continuum flux in every optical/UV band, and then correct the photometric data of the \xmm/OM filters to remove the emission line flux. We calculate the line flux contribution to each filter by convolving the spectra with the full 
response files of each optical/UV filters, read from the response files stored in the latest calibration database at \xmm\footnote{https://heasarc.gsfc.nasa.gov/FTP/xmm/data/responses/om}. We calculated the correction factor for every OM and UVOT filter, and then apply it to the corresponding photometric data. The typical correction factor is 5 -- 10 per cent, increasing from UV to optical. Finally these data are used as inputs for the SED fitting.

\subsubsection{The Accretion Disc Model}
\rxj0134\ is an extreme super-Eddington NLS1
(\citealt{Grupe.2010}). In the inner region of such a super-Eddington accretion flow, energy advection and/or disc winds can take away a significant amount of the accretion energy (e.g. \citealt{Poutanen.2007, Hagino.2016, Done.2016, Jin.2017b}). 
There are several AGN SED models in {\sc xspec} (\citealt{Arnaud.1996}), such as {\tt optxagnf} (\citealt{Done.2012}), {\tt agnsed} (\citealt{Kubota.2018}) and {\tt agnslim} (\citealt{Kubota.2019}). Only the latter adopts a slim disc emissivity, where the surface luminosity is kept at the local Eddington limit within a critical radius (\citealt{Abramowicz.1988, Watarai.2000, Wang.2003, Sadowski.2011, Kubota.2019}). {\tt agnslim} uniquely combines this maximum emissivity with the ability to change 
the local emission to change between blackbody, soft Comptonisation and hard Comptonisation, in order to model the disc, soft excess and hard X-ray corona emission, respectively. 

Another major difference between a standard disc and a slim disc is that the 
inner radius of the disc ($R_{\rm in}$) is 
determined by the gas pressure more than by the black hole spin for highly super-Eddington discs (e.g \citealt{Watarai.2000}). This removes the most obvious signature of black hole spin, so we conservatively fix spin at zero here, with the consequence of minimizing the inferred Eddington ratio of the flow.

By default, in the {\tt agnslim} model the seed photon temperature of the hard X-ray Comptonisation is set to be the temperature of inner disc photons. However, recent studies of X-ray {\it simple} super-Eddington NLS1s show that their X-ray spectral-timing properties are better modelled if the warm corona, rather than the inner disc, provides seed photons for the hot corona (\citealt{Jin.2013, Jin.2016, Jin.2017a, Jin.2021}). Thus we made a small modification to {\tt agnslim} to link the seed-photon temperature of the hard X-ray Comptonisation to the electron temperature of the warm corona.  We refer to this modified {\tt agnslim} model as {\tt agnslimhot}, and use it in our subsequent SED analysis. 

{\tt agnslimhot} inherits the full set of parameters of {\tt agnslim} (see Table~\ref{tab-sedfit}). The black 
hole mass $M_{\rm BH}$ is fixed at $2\times 10^7 M_{\odot}$. The comoving distance is fixed at 937.1 Mpc for redshift $z=0.237$. The inclination angle $\theta_{\rm inc}$ is fixed at 60\degr, which is larger than 30\degr\ as often assumed for normal X-ray {\it simple} super-Eddington NLS1s. This is because the enigmatic X-ray variability of \rxj0134\ implies complex and variable absorption, which is more likely to happen at a larger inclination angle. The electron temperature of the hot corona $kT_{\rm e, hot}$ is fixed at 200 keV, and the overall normalization is fixed at 1. The remaining free parameters include the electron temperature $kT_{\rm e, warm}$, photon index $\Gamma_{\rm warm}$ and radius $R_{\rm warm}$ of the warm corona; the photon index $\Gamma_{\rm hot}$ and radius $R_{\rm hot}$ of the hot corona; the mass accretion rate through the outer disc $\dot{m}_{\rm out}$ and the outer radius of the disc $R_{\rm out}$.

\begin{figure}
\centering
\begin{tabular}{c}
\includegraphics[trim=0.08in 0.3in 0.0in 0.0in, clip=1, scale=0.48]{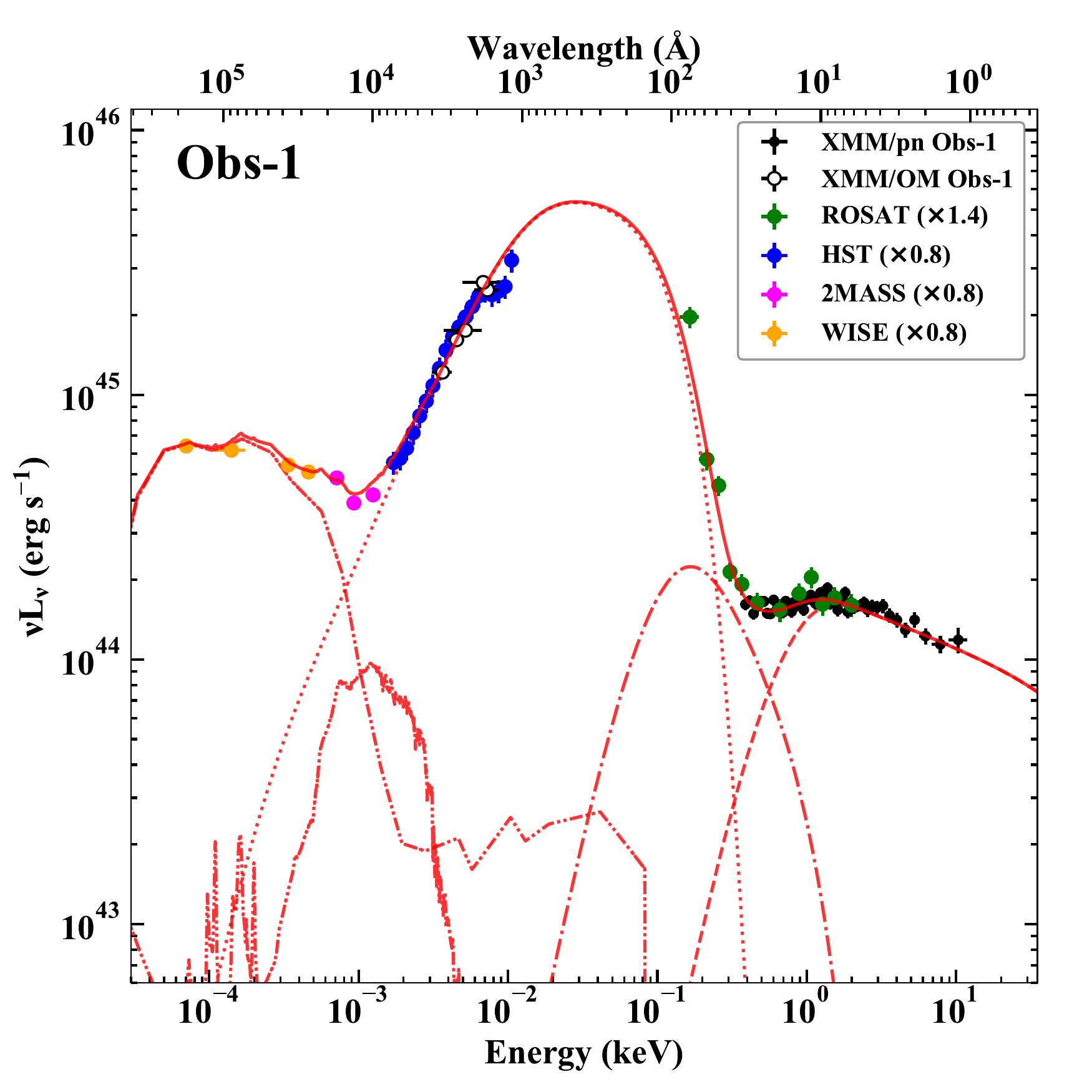}  \\
\end{tabular}
\caption{The unabsorbed and de-redshifted best-fit broadband SEDs of \rxj0134\ in Obs-1. 
The red solid line is the total best-fit model, which includes the hot dust emission (red dash-dot-dot line), host galaxy emission (red dash-dot-dot-dot line), accretion disc emission (dotted line), warm Comptonisation (red dash-dot line) and hot Comptonisation (red dash line). The scaling factors shown in the legend have been applied to different datasets to account for their normalization differences.}
\label{fig-sedfit}
\end{figure}

\subsubsection{Additional Components in the Broadband SED}
\rxj0134\ was originally classified as a radio-loud (RL) AGN with a radio loudness of $R=71$ (\citealt{Grupe.2000}), thus there is possibility that the X-ray emission might also include some contribution from the jet, such as the synchrotron self-Compton (SSC) and external Compton (EC) emission (e.g. \citealt{Kynoch.2018}). However, we did not find any evidence of jet emission from X-ray spectral-timing analysis (see Paper-I), and our ongoing radio/optical monitoring campaign shows that \rxj0134\ has returned to a radio-quiet state (see Paper-III). Therefore, the spectral components in the {\tt agnslimhot} model should be enough to fit the nuclear emission. 

To model the hot dust emission in the near infrared, we take the hot dust template from \citet{Silva.2004}, and import it into {\sc xspec} as a local {\tt zagndust} model. The host galaxy is not resolved in optical images, but it is still possible to identify its flux contribution in the spectrum. We assume it is an Sb galaxy similar to the famous NLS1 \rej1034, and adopt a corresponding galaxy spectral template from \citet{Polletta.2007}, which is loaded into {\sc xspec} as the local {\tt hostgal} model. Since not all the datasets are simultaneous or observed by the same instrument, there can be normalization discrepancies caused by e.g. long-term variability, different aperture size and flux calibration. Thus we use a free constant to account for the normalization differences between the {\it ROSAT}, \xmm\ and {\it HST} data. The data points from {\it 2MASS}, {\it WISE} and {\it HST} join smoothly with each other, and so we use the same constant for these three datasets.

This combination of accretion flow, host galaxy and hot dust, describe the intrinsic continuum, but these spectra are further modified by absorption and reddening along the line of sight. 

We use {\tt tbabs}/{\tt ztbabs} (\citealt{Wilms.2000}) to model the gas absorption from the Milky Way and host galaxy, respectively, with $N_{\rm H,gal}$ fixed at 1.77$\times10^{20}$ cm$^{-2}$ for our line of sight (\citealt{Willingale.2013}), and $N_{\rm H,host}$ left free. The absorption cross-sections were set to the values of \citet{Verner.1996}. 
While this model is a good approximation to the X-ray absorption, it is less good at modelling the impact of this same gas in the UV due to its assumption that the material is completely neutral (but the interstellar medium is multiphase, e.g. \citealt{McKee.1995, Wolfire.2003}) and that the UV absorption is dominated by bound-free edges rather than lines. 
Nonetheless, both the Galactic column and the host-galaxy column here are rather small (see Section~\ref{sec-sedfit1} below), so this mis-modelling of its UV absorption is not very important.

We also use {\tt redden}/{\tt zredden} to model the dust reddening associated with the gas in the Milky Way and the host galaxy, respectively. The Galactic reddening $E(B-V)_{\rm gal}$ is fixed at 0.0144 (\citealt{Schlegel.1998}), while
$E(B-V)_{\rm host}$ is left as a free parameter.
We note that {\tt zredden} may not be appropriate to describe the effect of dust in the host galaxy if 
this is associated with the nuclear region rather than in the interstellar medium (see e.g. \citealt{Collinson.2015}), but this has little impact here as the UV is clearly a very blue spectrum, so the reddening is most probably limited.

\begin{table}
\centering
\caption{The best-fit SED parameters of \rxj0134\ in Obs-1. The errors indicate 90 per cent confidence limits. `fixed' indicates that the parameter is fixed at the given value. $C_{\rm ROSAT}$ and $C_{\rm HST}$ are the scaling factors for the {\it ROSAT} and {\it HST} data.}
\begin{tabular}{llcc}
\hline
Component & Parameter & Value & Unit \\
\hline
{\tt tbabs} & $N_{\rm H, gal}$ & 1.77 fixed & $10^{20}$ cm$^{-2}$ \\
{\tt redden} & $E(B-V)_{\rm gal}$ & 1.44  fixed & $10^{-2}$ \\
{\tt ztbabs} & $N_{\rm H, host}$ & 0.56 $^{+0.43}_{-0.27}$ & $10^{20}$ cm$^{-2}$ \\
{\tt zredden} & $E(B-V)_{\rm host}$ & 0.40 $^{+0.27}_{-0.24}$ & $10^{-2}$ \\
{\tt zagndust} & norm & 1.90 $^{+0.09}_{-0.08}$ & $10^{-5}$ \\
{\tt hostgal} & norm & 1.05 $^{+0.30}_{-0.28}$  & $10^{-2}$ \\
{\tt agnslimhot} & $M_{\rm BH}$ & 2.0 fixed & $10^{7}M_{\odot}$ \\
{\tt agnslimhot} & log($\dot{m}_{\rm out}$) & 1.31 $^{+0.01}_{-0.01}$ & \\
{\tt agnslimhot} & $a_{*}$ & 0.0 fixed  & \\
{\tt agnslimhot} & cos $\theta_{\rm inc}$ & 0.5 fixed \\
{\tt agnslimhot} & $kT_{\rm e, warm}$ & 0.22 $^{+0.37}_{-0.08}$ & keV \\
{\tt agnslimhot} & $kT_{\rm e, hot}$ & 200 fixed & keV \\
{\tt agnslimhot} & $\Gamma_{\rm hot}$ & 2.22 $^{+0.05}_{-0.04}$ & \\
{\tt agnslimhot} & $\Gamma_{\rm warm}$ & 2.84 $^{+1.46}_{-1.62}$ &  \\
{\tt agnslimhot} & $R_{\rm hot}$ & 4.65 $^{+0.13}_{-0.16}$    & $R_{\rm g}$ \\
{\tt agnslimhot} & $R_{\rm warm}$ & 5.44 $^{+0.87}_{-0.28}$    & $R_{\rm g}$ \\
{\tt agnslimhot} & log($R_{\rm out}$) & 4.97 $^{+0.24}_{-0.22}$  & $R_{\rm g}$ \\
{\tt $C_{\rm ROSAT}$} &  & 0.75 $^{+0.04}_{-0.05}$ & \\
{\tt $C_{\rm HST}$} &  & 1.16 $^{+0.05}_{-0.05}$  & \\
\multicolumn{2}{l}{$\chi^{2}_{\nu}$}  & 607.3/475  & \\
\hline
\end{tabular}
\label{tab-sedfit}
\end{table}

\subsection{Results of the SED Modelling}
\label{sec-sedfit1}
Based on the above datasets and model configurations, we obtain the best-fit broadband SED for \rxj0134. 
Figure~\ref{fig-sedfit}a shows this
SED model, where both model and data are corrected for the Galactic and intrinsic extinction/absorption, and shown in the AGN rest-frame. The best-fit parameters are listed in Table~\ref{tab-sedfit}.

It is clear that the near-IR emission is dominated by the hot dust components, while the
UV continuum is well-fitted by the accretion disc component. There is a small ($\sim 10$ per cent) 
contribution from host galaxy star light
between these two in the optical/near-IR band. 
At higher energies, the 
soft X-ray emission observed by {\it ROSAT} below 0.3 keV is dominated by the emission from the inner disc. The hard X-rays above 2 keV are dominated by a hot corona with photon index of  2.22$^{+0.05}_{-0.04}$. There is some evidence for a warm Comptonisation component, with electron temperature of 0.22$^{+0.37}_{-0.08}$ keV and photon index 2.84$^{+1.46}_{-1.62}$. All these Comptonisation parameters are typical for X-ray {\it simple} super-Eddington NLS1s (e.g. \citealt{Jin.2013, Jin.2016, Jin.2017a}), apart from the soft X-ray excess being much weaker relative to the disc and hot corona as discussed in Paper-I
(see also the explicit comparison to \rx04\ in Section \ref{sec-wls}). 

The mass accretion rate through the outer disc ($\dot{m}_{\rm out}$) is completely determined by the observed optical/UV emission for the fixed black hole mass and spin, giving $\dot{m}_{\rm out}=20.6^{+0.3}_{-0.6}$, confirming that the accretion flow is highly super-Eddington. A higher value of black hole spin will only increase this. The only way to significantly reduce $\dot{m}_{\rm out}$ is to go to higher black hole mass, as the monochromatic luminosity on the Rayleigh-Jeans part of the standard (multi-temperature blackbody) disc continuum has $L_{\rm\nu} \propto (M_{\rm BH}~\dot{M})^{2/3} \propto (M_{\rm BH}^2~\dot{m}_{\rm out})^{2/3}$, where $\dot{m}_{\rm out}=\dot{M}/\dot{M}_{\rm Edd}$ (e.g. \citealt{Shakura.1973, Davis.2011, Kubota.2019}). Increasing the mass by a factor of 2 (i.e. $4\times 10^7 M_\odot$) then reduces $\dot{m}_{\rm out}$ by a factor of 4, but then $\dot{m}_{\rm out}$ is still $\sim 5$, so the disc is still super-Eddington even a factor of 2 away from our preferred black hole mass. 

The observed bolometric luminosity is derived by integrating the best fit model, and gives $L_{\rm bol}=1.63\times 10^{46}$~ergs s$^{-1}$. Thus $L_{\rm bol}/L_{\rm Edd}=6.3$, substantially below the $\dot{m}_{\rm out}=20.6$ derived for the accretion flow itself. This is clear evidence for a loss of power through advection and/or winds, as expected for a strongly super-Eddington flow.

\begin{table*}
\centering
   \caption{Comparison of optical/UV emission line properties among super-Eddington NLS1s, WLQs and the composite QSO spectra from \citet{Francis.1991} (F19). The line fitting method is described in Appendix~\ref{sec-optuvfit}. The reported equivalent widths are for the emission lines of Ly$\alpha$ plus the N {\sc v} $\lambda$1238/1243 doublet, C {\sc iv} $\lambda$1548/1551 doublet, Si {\sc iv} $\lambda$1393/1402 doublet, Mg {\sc ii} $\lambda$2797/2803 doublet, Fe {\sc ii} (4434 -- 4684 \AA) and [O {\sc iii}] $\lambda$5007. $v_{\rm blue}$ is the velocity of the blue-shifted Gaussian component in that emission line. $R_{\rm FeII}$ is the flux ratio between the Fe {\sc ii} (4434 -- 4684 \AA) and H$\beta$ broad Gaussian component.}
     \begin{tabular*}{\textwidth}{@{}lcccccccccc@{}}
     \hline
      Source &  Ly$\alpha$ + N {\sc v} REW & Si {\sc iv} REW & C {\sc iv} REW & C {\sc iv} $v_{\rm blue}$ & Mg {\sc ii} REW  & Fe {\sc ii} REW & H$\beta$ REW &  H$\beta$ FWHM & [O {\sc iii}] REW & $R_{\rm FeII}$ \\
      & (\AA) & (\AA) & (\AA) & (km s$^{-1}$) & (\AA) & (\AA) & (\AA) & (km s$^{-1}$) & (\AA) & \\
     \hline
     RE10 & 85.6 $\pm$ 8.6 & 22.6 $\pm$ 3.2 & 62.1 $\pm$ 5.3 & -620 $\pm$ 550 & 22.6 $\pm$ 1.6 & 24.4 $\pm$ 1.8 & 31.7 $\pm$ 1.4 & 620 $\pm$ 20 & 33.1 $\pm$ 2.3 & 0.79 $\pm$ 0.07  \\
     \ph1092 & 23.5 $\pm$ 2.4 & 6.6 $\pm$ 2.1 & 12.1 $\pm$ 2.5 & -9900 $\pm$ 2400 & -- & 53.2 $\pm$ 4.2 & 26.1 $\pm$ 5.4 & 1700 $\pm$ 300 & 4.5 $\pm$ 1.3 & 2.33 $\pm$ 0.45  \\
     \phl1811 & 28.6 $\pm$ 2.9 & 7.9 $\pm$ 2.0 & 7.9 $\pm$ 2.5 & -1900 $\pm$ 1200 & 13.8 $\pm$ 2.0 & 33.6 $\pm$ 3.0 & 36.1 $\pm$ 3.9  & 1900 $\pm$ 200 & 3.3 $\pm$ 2.8 & 1.11 $\pm$ 0.15 \\
     RX04 & 37.3 $\pm$ 3.7 & 7.5 $\pm$ 2.0 & -- & -- & -- & 27.6 $\pm$ 2.0 & 24.3 $\pm$ 2.5  & 4000 $\pm$ 800 & 6.3 $\pm$ 0.7 & 1.34 $\pm$ 0.16 \\
     1H07 & 39.1 $\pm$ 3.9 & 6.9 $\pm$ 4.5 & 21.6 $\pm$ 5.9 & -1200 $\pm$ 1100 & 9.8 $\pm$ 1.3 & 44.1 $\pm$ 3.3 & 25.4 $\pm$ 7.0  & 680 $\pm$ 60 & 4.7 $\pm$ 3.0 & 2.01 $\pm$ 0.47 \\
     RX01 & 24.4 $\pm$ 2.4 & 8.4 $\pm$ 1.2 & 7.6 $\pm$ 1.5 & -7600 $\pm$ 5700 & 12.1 $\pm$ 2.1 & 41.5 $\pm$ 3.4 & 23.8 $\pm$ 1.2  & 1410 $\pm$ 70 & 2.7 $\pm$ 0.8 & 1.74 $\pm$ 0.16 \\
     (F19) & 49.7 $\pm$ 5.0 & 9.0 $\pm$ 4.2 & 32.6 $\pm$ 7.0 & -830 $\pm$ 510 & 26.7 $\pm$ 2.7 & 31.2 $\pm$ 9.3 & 67.5 $\pm$ 4.1  & 3210 $\pm$ 40 & 17.9 $\pm$ 8.2 & 0.53 $\pm$ 0.15 \\
     \hline
     \end{tabular*}
\label{tab-comp-line}
\end{table*}

\begin{figure*}
\centering
\begin{tabular}{cc}
\includegraphics[trim=0.2in 0.4in 0.0in 0.0in, clip=1, scale=0.47]{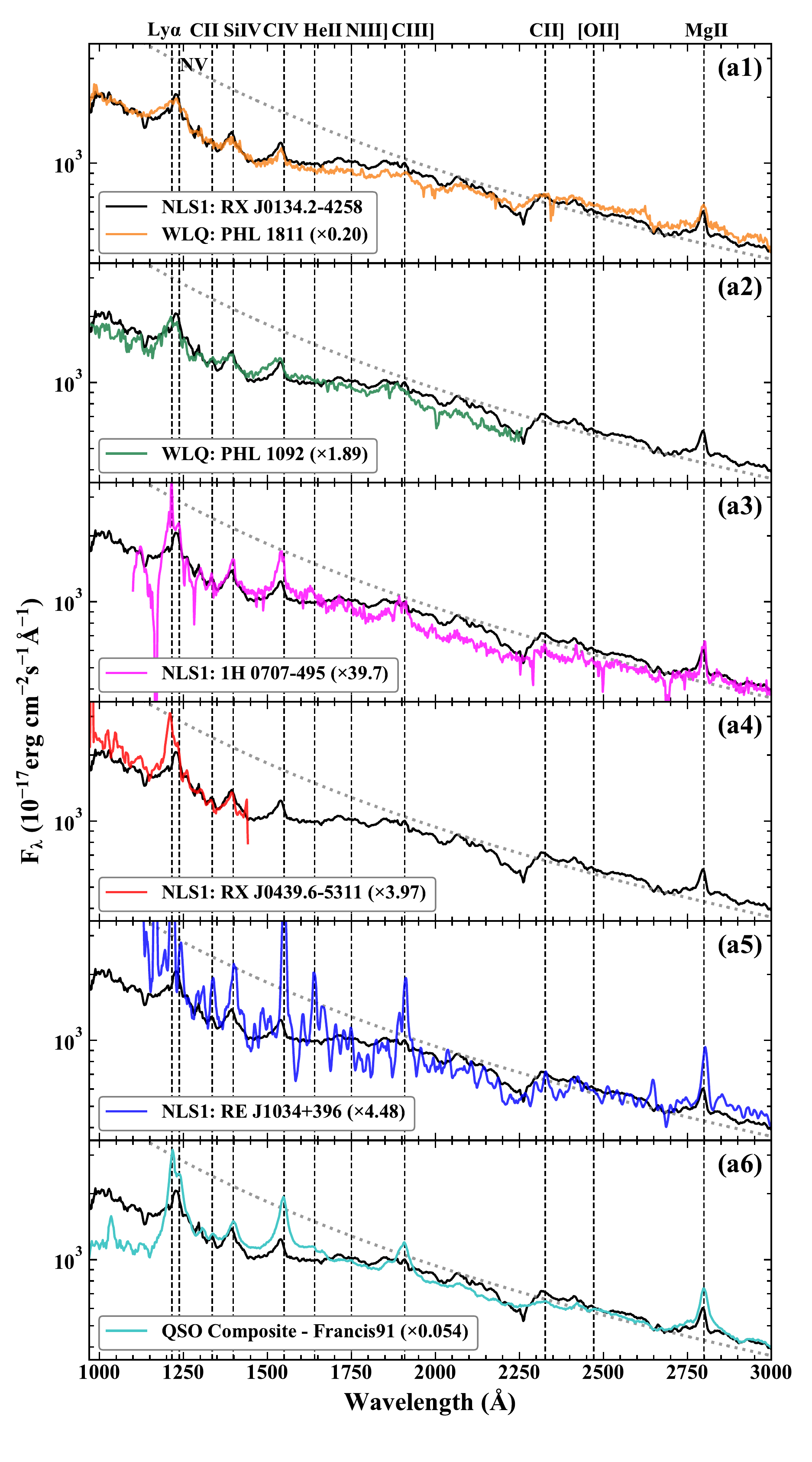} &
\includegraphics[trim=0.3in 0.4in 0.0in 0.0in, clip=1, scale=0.47]{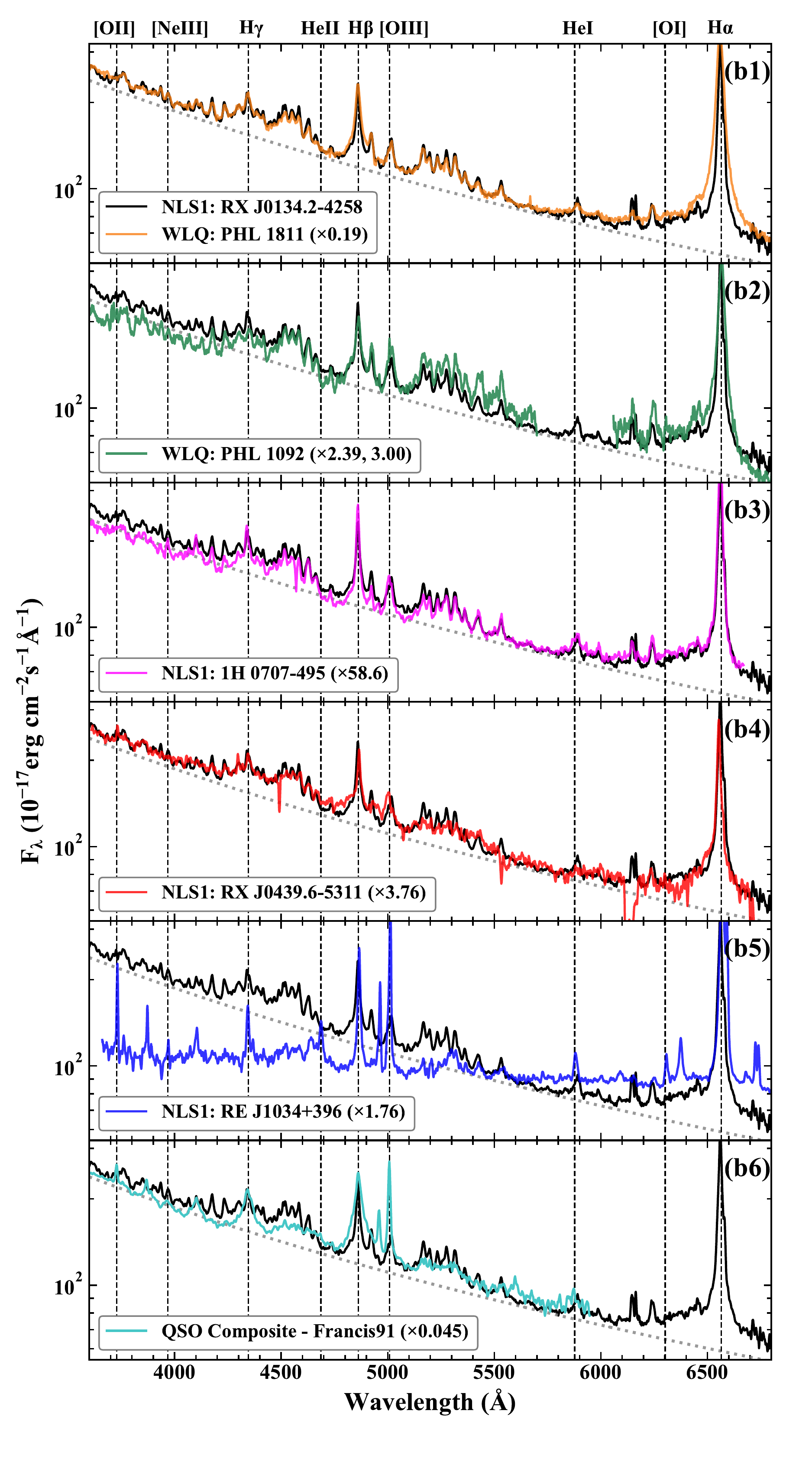} \\
\end{tabular}
\caption{Comparing the optical/UV spectrum of \rxj0134\ (black) with the WLQ PHL 1811 (orange) and PHL 1092 (green), the X-ray {\it complex} super-Eddington NLS1 1H 0707-495 (magenta), the X-ray {\it simple} super-Eddington NLS1 RX J0439.6-5311 (red), the X-ray QPO NLS1 RE J1034+396 (blue) and a QSO composite spectrum (light blue, from \citealt{Francis.1991}). In every panel, the standard thin disc model with flux density $\propto\lambda^{-7/3}$ is plotted as the dotted grey curve for comparison. All the spectra have been corrected for the Galactic reddening, de-redshifted to their rest-frames and rescaled to the similar flux level of \rxj0134\ for ease of comparison. The scaling factor is shown in every panel. There are two separate optical spectral segments for PHL 1092, thus different scaling factors are applied. The UV spectra of all the sources are all from archival {\it HST} observations. References for the optical spectra are as follows: RX J0134.2-4258 (this work), PHL 1811 (\citealt{Leighly.2007b}), PHL 1092 (\citealt{Marinello.2020}), 1H 0707-495 (\citealt{Done.2016}), RX J0439.6-5311 (\citealt{Jin.2017b}) and RE J1034+396 (\citealt{Jin.2012a}). }
\label{fig-optuv-compare}
\end{figure*}

\begin{table*}
\centering
   \caption{Comparison of the broadband SED properties among super-Eddington NLS1s, WLQs and mean quasar properties. For individual sources, typical black hole masses are adopted  from the literatures listed below. Other parameters are measured from the SEDs shown in Figure~\ref{fig-sed-compare}. $L_{\rm bol}$ and $\dot{m}_{\rm out}$ are the bolometric luminosity and mass accretion rate through the outer disc. We do not provide errors for  $L_{\rm bol}$, $\dot{m}$ or related parameters as they should be dominated by systematic SED model uncertainties which are difficult to estimate. $\mu$ is the observed radiative efficiency. $\mu_{0}=0.057$ is for the standard \citet{Shakura.1973} disc with zero spin. $k_{\rm 2-10keV}$, $k_{\rm 0.5-1keV}$ and $k_{5100\textup{\AA}}$ are the bolometric corrections for 2-10 keV, 0.5-1 keV and 5100 \AA. $f_{\rm IR}$ is the fraction of infrared dust luminosity in 1-50 $\mu$m relative to $L_{\rm bol}$. $\alpha_{\rm ox}$ and $\alpha_{\rm optir}$ are the optical-to-X-ray and optical-to-infrared spectral indices. Their statistical 1-$\sigma$ errors are provided. The relative uncertainty of infrared luminosity is assumed to be 10 per cent. Zero spin is assumed for all the parameter values listed in this table.}
     \begin{tabular}{ccccccccccccc}
     \hline
      & $M_{\rm BH}$ & $L_{\rm bol}$ & $L_{\rm bol}/L_{\rm Edd}$ & $\dot{m}_{\rm out}$ & $\mu$ & $\mu/\mu_{\rm 0}$ & $k_{\rm 2-10keV}$  & $k_{\rm 0.5-1keV}$ & $k_{\rm 5100\textup{\AA}}$ & $\alpha_{\rm ox}$ & $\alpha_{\rm optir}$ & $f_{\rm dust}$ \\
     & ($10^{7}M_{\odot}$) & ($10^{45}$ erg s$^{-1}$) & & & & & & & & & & (\%)  \\
     \hline
     \multicolumn{13}{@{}l}{RE J1034+396: (\rej1034, $z=0.042$, moderate-$\dot{m}$ X-ray {\it simple} NLS1)} \\
     Obs-1  & 0.2 & 0.5 & 1.7 & 1.7 & 0.057 & 1.00 & 226 & 64 & 60 & 1.33 $\pm$ 0.07 & 1.23 $\pm$ 0.15 & 16  \\
     \hline
     \multicolumn{13}{@{}l}{PHL 1092 (\ph1092, $z=0.396$, WLQ)} \\
     Obs-1  & 10.0  & 29.8 & 2.3 & 2.3 & 0.057 & 1.00 & 399 & 61 & 23 & 1.69 $\pm$ 0.07 & 0.94 $\pm$ 0.08 & 26  \\
     Obs-2  & 10.0  & 29.1 & 2.2 & 2.2 & 0.057 & 1.00 & 56900 & 12700 & 23 & 2.49 $\pm$ 0.25 & 0.87 $\pm$ 0.08 & 22  \\
     \hline
      \multicolumn{13}{@{}l}{PHL 1811 (\phl1811, $z=0.192$, WLQ)} \\
     Obs-1 & 15.0  & 89.2 & 4.6 & 7.4 & 0.035 & 0.62 & 16500 & 37900 & 17 & 2.46 $\pm$ 0.11 & 0.67 $\pm$ 0.08 & 17  \\
     \hline
      \multicolumn{13}{@{}l}{RX J0439.6-5311 (\rx04, $z=0.242$, high-$\dot{m}$ NLS1: X-ray {\it simple})} \\
     Obs-1     & 0.7 & 5.2 & 5.7 & 11.3 & 0.029 & 0.50 & 90 & 10 & 24 & 1.22 $\pm$ 0.04 & 0.91 $\pm$ 0.11 & 11 \\
     \hline
      \multicolumn{13}{@{}l}{1H 0707-495 (\1h07, $z=0.041$, high-$\dot{m}$ NLS1: X-ray {\it complex})} \\
     Obs-1        & 0.2  & 2.1 & 8.0 & 22.7 & 0.020 & 0.35 & 320 & 69 & 44 & 1.46 $\pm$ 0.04 & 0.57 $\pm$ 0.09 & 4  \\
     Obs-2        & 0.2  & 2.6 & 9.9 & 31.2 & 0.018 & 0.32 & 1270 & 340 & 44 & 1.85 $\pm$ 0.06 & 0.45 $\pm$ 0.09 & 3  \\
     Obs-3        & 0.2  & 1.7 & 6.5 & 26.4 & 0.014 & 0.25 & 2100 & 1460 & 33 & 2.02 $\pm$ 0.08 & 0.55 $\pm$ 0.09 & 5  \\
     \hline
     \multicolumn{13}{@{}l}{RX J0134.2-4258 (\rxj0134, $z=0.237$, high-$\dot{m}$ NLS1 \& WLS)}\\
     Obs-1    & 2.0  & 16.3 & 6.3 & 20.6 & 0.017 & 0.31 & 75 & 149 & 22 & 1.40 $\pm$ 0.02 & 0.66 $\pm$ 0.08 & 12  \\
     Obs-2    & 2.0  & 17.3 & 6.7 & 26.2 & 0.015 & 0.26 & 319 & 600 & 20 & 1.70 $\pm$ 0.02 & 0.66 $\pm$ 0.08 & 13  \\
     \hline
     \multicolumn{13}{@{}l}{Quasar Mean SED (\citealt{Elvis.1994})} \\
     -- & -- & -- & -- & -- & -- & -- & 27 & 48 & 6 & 1.38  & 0.80  & 34  \\
     \hline
     \multicolumn{13}{@{}l}{Quasar Mean SED (\citealt{Richards.2006})} \\
     -- & -- & -- & -- & -- & -- & -- & 103 & 109 & 4 & 1.53 & 0.85  & 37  \\
     \hline
     \end{tabular}
     \\References for the redshifts and black hole mass estimates: \rej1034: \citealt{Jin.2021}; \ph1092: \citealt{Miniutti.2012, Marinello.2020}; \phl1811: \citealt{Leighly.2007a, Leighly.2007b}; \rx04: \citealt{Jin.2017a}, \citealt{Jin.2017b}; \1h07: \citealt{Done.2016}; \rxj0134: this work; Quasar mean SEDs: \citealt{Elvis.1994}, \citealt{Richards.2006}.
\label{tab-comp-sed}
\end{table*}

\begin{figure*}
\centering
\begin{tabular}{cc}
\includegraphics[trim=0.1in 0.in 0.0in 0.0in, clip=1, scale=0.48]{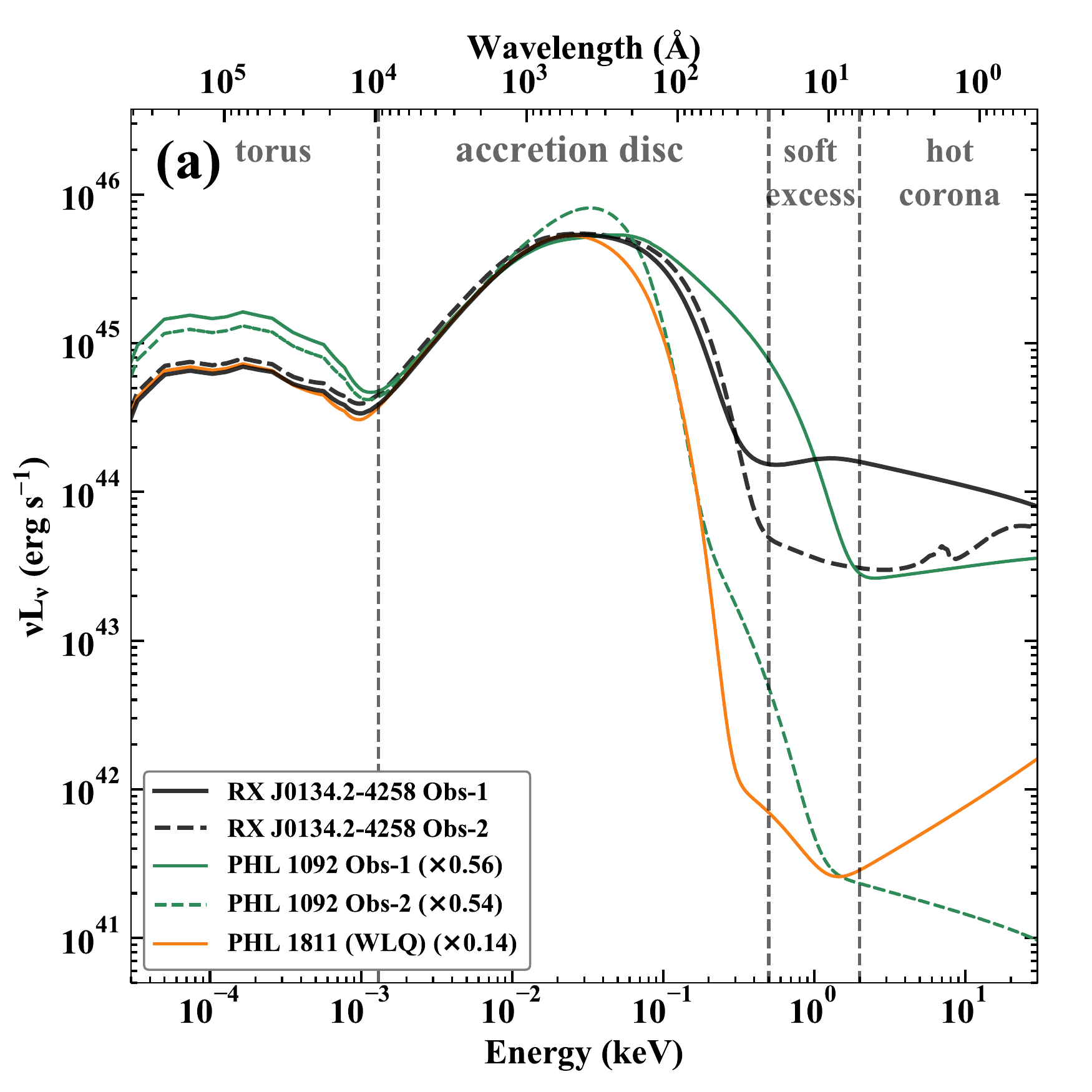} &
\includegraphics[trim=0.2in 0.in 0.0in 0.0in, clip=1, scale=0.48]{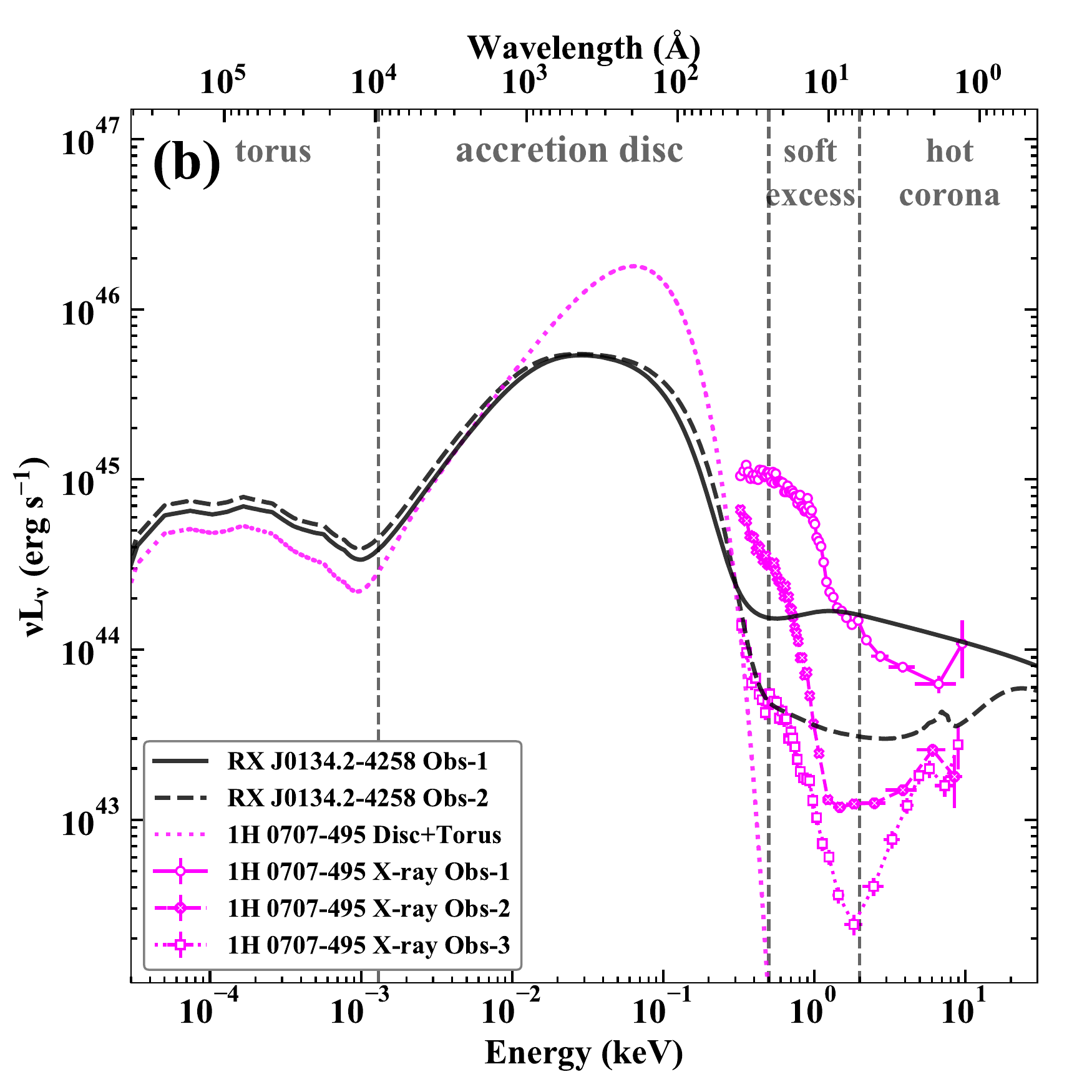} \\
\includegraphics[trim=0.1in 0.in 0.0in 0.0in, clip=1, scale=0.48]{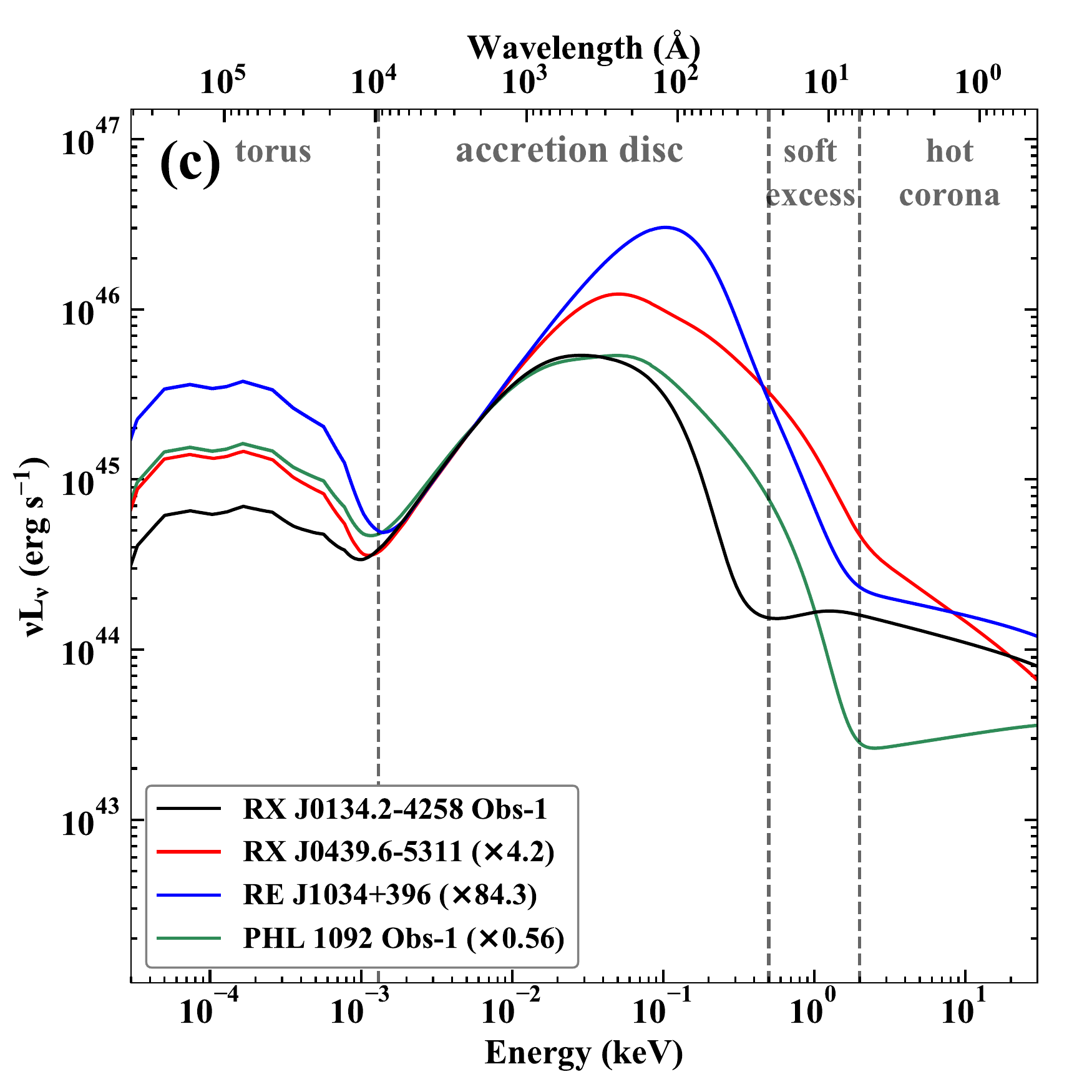} &
\includegraphics[trim=0.2in 0.in 0.0in 0.0in, clip=1, scale=0.48]{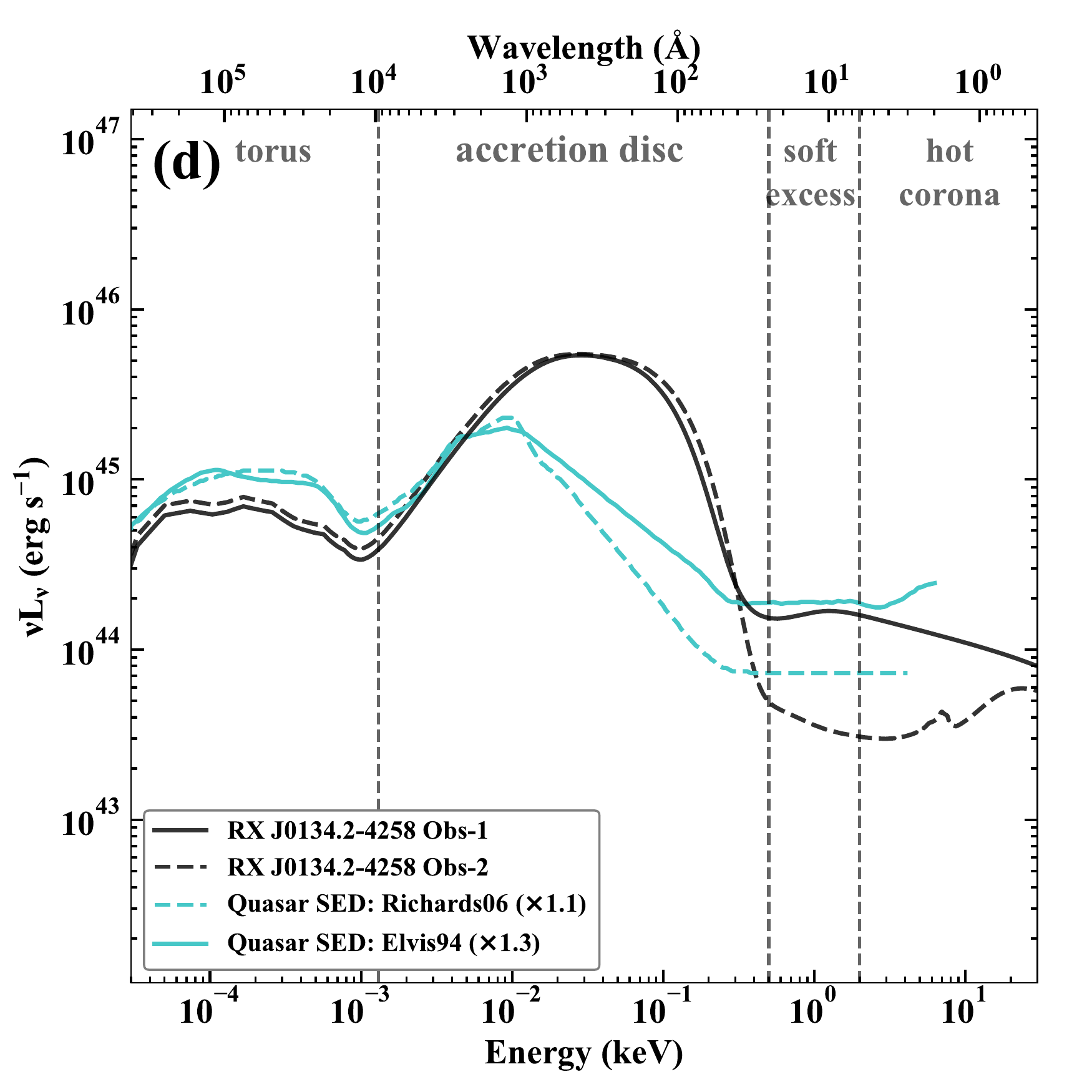} \\
\end{tabular}
\caption{
Comparing the two SEDs of \rxj0134\ in its Obs-1 and Obs-2 with some other AGN, whose SEDs have been rescaled to match the SED of \rxj0134\ in Obs-1 at 2500 \AA. The scaling factors have been shown in every panel. Panel-a: comparing the two SEDs of \rxj0134\ (Obs-1: black solid, Obs-2: black dash) with two WLQs, namely \phl1811\ (orange) and \ph1092\ (green). The high-flux SED of \ph1092\ was observed by \xmm\ in 2003-07-18 (Obs-1, green solid), while the low-flux SED was observed in 2008-01-20 (Obs-1, green dash). Panel-b: comparing with the representative X-ray {\it complex} super-Eddington NLS1 \1h07\ (\citealt{Done.2016}). We show three distinct X-ray spectra from different \xmm\ observations in 2010-09-15 (Obs-1, magenta), 2007-05-14 (Obs-2, green) and 2011-01-12 (Obs-3, blue). The best-fit disc plus torus model of \1h07\  (magenta dotted line) is based on the OM data obtained in its Obs-1. Panel-c: comparing with two representative X-ray {\it simple} super-Eddington NLS1s, including \rx04\ (\citealt{Jin.2017b}) and \rej1034\ (\citealt{Jin.2021}). The Obs-1 SED of \ph1092\ (green solid) is also plotted for comparison. Panel-d: comparing with the quasar mean SEDs reported by \citet{Elvis.1994} (cyan solid) and \citet{Richards.2006} (cyan dash).
}
\label{fig-sed-compare}
\end{figure*}

\section{Discussion}
\subsection{Weak UV Lines and the SEDs of Super-Eddington Flows}
\label{sec-wls}
The weak UV lines of \src1\ resemble those from WLQs, defined as Ly$\alpha$ + N {\sc v} REW $\le$ 15 \AA, and C {\sc iv} REW $\le$ 10 \AA\ (\citealt{Fan.1999, Diamond-Stanic.2009, Plotkin.2010, Wu.2011, Wu.2012, Luo.2015, Ni.2018}). In this section, we compare \src1\ with some typical WLQs and NLS1s (see Table~\ref{tab-comp-sed})\footnote{Note that the estimates of the mass, mass accretion rate and bolometric luminosity are all subject to various systematic uncertainties, so we only show their typical values and do not provide their uncertainties in Table \ref{tab-comp-sed}.}, in order to obtain a full understanding of their similarities and differences.

\subsubsection{PHL 1811 (hereafter: \phl1811): typical WLQ}
So far, most of the known WLQs lie at relatively high redshifts with $z>2.2$. This redshift distribution is partly a selection effect as the C {\sc iv} line is redshifted into the more easily accessible optical band where wide area surveys
such as Sloan Digital Sky Survey (SDSS, \citealt{York.2000}, 3000 -- 9200 \AA) detect a large number of sources. 
However, it is also possible to find WLQs at lower redshifts from pointed UV spectroscopic observations (e.g. \citealt{McDowell.1995, Londish.2004}), such as the classic WLQ \phl1811\ at $z=0.192$, whose black hole mass is $\sim1.5\times10^{8}M_{\odot}$ and mass accretion rate $\sim {7}$ (\citealt{Leighly.2007a, Leighly.2007b}; \citealt{Wu.2012, Luo.2015}).

\citet{Leighly.2007a} first noticed the similarity of weak UV lines between \phl1811\ and \src1\ in the wavelength range of 1000-1600\AA. Here we compare the entire optical/UV spectra of \src1\ (black) and \phl1811\ (orange), as shown in Figure~\ref{fig-optuv-compare} panels a1 and b1. It is clear that these two AGN have remarkably similar optical/UV lines, including the strong optical Fe {\sc II} lines, extremely weak oxygen forbidden lines, and very weak and blue-shifted UV C {\sc iv}, C {\sc iii}] and N {\sc v} lines. Their optical/UV underlying continua also have similar shapes. Their optical continua, after correcting for the Galactic reddening, are consistent with a standard thin disc model (dotted grey curve: \citealt{Shakura.1973}), while their UV continua appear flatter at $\lesssim$ 2300 \AA. We apply the same line fitting method as described in Section~\ref{sec-uv-spec} to measure the REWs of the UV lines of \phl1811. We find Ly$\alpha$ + N {\sc v} REW = 28.6 $\pm$ 2.9 \AA\ and C {\sc iv} REW = 7.9 $\pm$ 2.0 \AA\ (for C {\sc iv} $\lambda$1548 + C {\sc iv} $\lambda$1551) for \phl1811, which are remarkably similar to \src1 (see Table~\ref{tab-comp-line}).

However, \phl1811\ has a black hole mass which is  around an order of magnitude higher than \src1. We fit the broad band SED of 
\phl1811\ with the same models as used for \src1, fixing black hole spin at $a_{*}=0$, so the model parameters are all directly comparable. This gives $\dot{m}_{\rm out}=7.4$ (see Table \ref{tab-comp-sed}), a factor of 3 lower than that of \src1.

We compare their broad band SEDs as the solid black and solid orange lines in Figure \ref{fig-sed-compare}a. The lower black hole mass and higher $\dot{m}_{\rm out}$ of \src1\ predict a higher inner disc temperature, but it is clear that the WLQ \phl1811\ has much less X-ray emission relative to the disc peak, and that the X-rays have complex shape. More quantitatively, we adopt the optical-to-X-ray index ($\alpha_{\rm ox}$, e.g. \citealt{Lusso.2010}) as,
\begin{equation}
\alpha_{\rm ox} = - \frac{\rm{log}(L_{\rm 2 keV}/L_{\rm 2500\textup{\AA}})}{\rm{log}(\nu_{\rm 2 keV}/\nu_{\rm 2500\textup{\AA}})}
\end{equation}
where $L_{\rm 2 keV}$ and $L_{\rm 2500\textup{\AA}}$ are the luminosities at 2 keV and 2500 \AA. We find $\alpha_{\rm ox}$ is 2.46 $\pm$ 0.11  for \phl1811, but is only 1.40 $\pm$ 0.02 for Obs-1 of \src1. Paper-I shows that \src1\ also has a complex X-ray spectral shape when it is X-ray weaker. We include this Obs-2 spectrum as the dashed black line in Figure~\ref{fig-sed-compare}a. Clearly it is not so extremely X-ray weak as \phl1811, as confirmed by its $\alpha_{\rm ox}$ of 1.70 $\pm$ 0.02, but the same mechanisms may well be at work, most likely absorption and scattering in a clumpy wind (\citealt{Done.2016, Jin.2017b}). Such winds are clearly expected from super-Eddington sources. 

Then we compare the broadband SED and optical/UV spectra of \src1\ to some other well-known super-Eddington AGN. 

\subsubsection{PHL 1092 (hereafter: \ph1092): typical WLQ}
\ph1092, lying at $z=0.396$, is often discussed as a WLQ, though its UV line REWs are slightly larger than the formal definition. Its mass is $M_{\rm BH}=1.0\times10^{8}M_{\odot}$, similar to \phl1811, but is one order of magnitude larger than \src1. Its mass accretion rate is only mildly super-Eddington at $\dot{m}_{\rm out}=2.2$ (\citealt{Miniutti.2012, Marinello.2020}, see Table \ref{tab-comp-sed}). Nonetheless, Figure~\ref{fig-optuv-compare} panels a2 and b2 shows its UV/optical spectra (green line) are very similar in both continuum shape and line emission to \src1\ (black line). 
Its broadband SED shows dramatic soft and hard X-ray variability. One SED observation (Obs-1 of \ph1092, Figure~\ref{fig-sed-compare}a, green solid line) shows a strong soft X-ray excess, together with harder X-ray emission which has $\alpha_{\rm ox}=1.69$, similar ratio to the disc power as in \src1. The other SED observation (Obs-2 of \ph1092, green dashed line) shows a much reduced soft X-ray flux, together with extremely weak and complex harder X-ray emission, whose $\alpha_{\rm ox}$ increases to 2.49 $\pm$ 0.25, similar to the X-ray weak WLQ \phl1811.

\subsubsection{1H 0707-495 (hereafter: \1h07): high-$\dot{m}$ NLS1}
A dramatic change in the soft and hard X-ray fluxes is also seen in the famous NLS1 \1h07. This source shows a strong soft X-ray excess and stochastic dips in the 0.3-10 keV light curve as observed by e.g. \xmm\ and {\it eROSITA} (\citealt{Wilkins.2014, Done.2016, Boller.2021}). The drastic X-ray variability may be originated from complex absorption in the wind (\citealt{Hagino.2016, Parker.2021}).
Besides, it has a similar super-Eddington accretion flow as \src1, with $\dot{m}\sim20-30$, by an order of magnitude lower mass, $M_{\rm BH}=2\times10^{6}~M_{\odot}$ (\citealt{Done.2016}, see table \ref{tab-comp-sed}).
We show its UV/optical spectra in magenta in Figure~\ref{fig-optuv-compare} panels a3 and b3. Plainly its optical continuum and lines are very similar to \src1\ and WLQs. However, the UV lines are now not so much alike, in that \1h07\ shows a substantial REW core which is not strongly blue-shifted (see table \ref{tab-comp-line}). Assuming that the blue-shifted wing of the C {\sc iv} is showing the wind strength, then \1h07\ has a less strong wind despite the accretion flow being similarly super-Eddington to \src1, and much more super-Eddington than \phl1811\ and \ph1092. Figure~\ref{fig-sed-compare}b shows three distinct SEDs of \1h07\ (magenta lines and data, similar to those reported by \citealt{Done.2016}) compared to both observations of \src1\ (black solid and dashed lines).
This clearly shows that not only are the hard X-rays affected by this complex absorption/scattering in the wind, but the shape and strength of the soft X-rays can also change. As a result, the optical-to-X-ray index $\alpha_{\rm ox}$ varies between 1.4 and 2.0. Thus it may also be that the strangely weak soft X-ray excess in \src1\ is due to some obscurers in our line-of-sight rather than being intrinsic.

\subsubsection{RX J0439.6-5311 (hereafter: \rx04): high-$\dot{m}$ NLS1}
We explore this further using another super-Eddington source, \rx04\ ($\dot{m}_{\rm out}=5.9$ for $M_{\rm BH}=7\times10^{6}~M_{\odot}$: \citealt{Jin.2017b}, see Table \ref{tab-comp-sed}). This is an archetypal X-ray {\it simple} NLS1 so has very little complex X-ray variability, making it likely that we have a clean line-of-sight to this object.
Figure~\ref{fig-optuv-compare} panels a4 and b4 shows the UV and optical spectra of \rx04\ (red spectra) compared to \src1. Again there is a continuum bend in the UV, indicating that this is an intrinsic feature rather than being due to reddening in the host galaxy. Unfortunately, the {\it HST} spectra do not cover the C {\sc iv} line, but the Ly$\alpha$+N {\sc v} REW is larger than the formal definition of a WLQ despite the similarity in Si {\sc iv} line shape and REW (see Table~\ref{tab-comp-line}).
Figure~\ref{fig-sed-compare}c shows the broadband SED of \rx04\ (red line) compared to the Obs-1 SED of \src1\ (black line). The soft X-ray excess of \rx04\ is much stronger than \src1, which is quantitatively confirmed by the smaller optical-to-X-ray index $\alpha_{\rm ox}$ of \rx04\ (see Table~\ref{tab-comp-sed}).

A possible explanation is that the soft excess of \src1\ is intrinsically present, but is severely suppressed due to complex and variable obscuration/scattering in the wind. Indeed, this is also seen in the variability of \ph1092\ and \1h07\ that the extent of the soft X-ray emission can be dramatically reduced. Especially, the soft X-ray bright state of \ph1092\ (its Obs-1, green solid line in Figure~\ref{fig-sed-compare}c) looks like \rx04, while the soft X-ray weak state of \ph1092\ is more like \src1. However, while wind obscuration/scattering seems to be a good explanation for the extremely weak soft excess of \src1, it remains difficult to understand how its hard X-rays are apparently seen directly, i.e. not suppressed by the wind as much as the soft excess does.

\subsubsection{RE J1034+396 (hereafter: \rej1034): moderate-$\dot{m}$ NLS1}
Alternatively, we explore the idea that the very steep soft X-ray emission of \src1\ is intrinsic. One of the steepest soft X-ray AGN is \rej1034, a source which uniquely shows a persistent X-ray quasi-periodic oscillation in AGN (QPO: \citealt{Gierlinski.2008}; \citealt{Middleton.2011}; \citealt{Alston.2014}; \citealt{Jin.2020}). This is a very low mass black hole, with $\dot{m}_{\rm out}=1.7$ for $M_{\rm BH}=2\times10^{6}~M_{\odot}$ (\citealt{Jin.2021}). 
Figure~\ref{fig-optuv-compare} panels a5 and b5 shows its UV and optical spectra (blue spectra)
compared to \src1. Plainly the optical spectrum is very different, probably due to strong host galaxy contamination (e.g. \citealt{Czerny.2016, Jin.2021}). The UV continuum shape is similar to \src1 (black line) but the UV lines are much stronger, making it unlike a WLQ (see Table~\ref{tab-comp-line}). Figure~\ref{fig-sed-compare}c also includes the broadband SED of \rej1034\ (blue line), with a similar hard X-ray shape as in \src1, but a much stronger steep soft X-ray excess. The optical-to-X-ray index $\alpha_{\rm ox}$ of \rej1034\ is 1.33 $\pm$ 0.07. The very steep soft X-ray emission here is most likely the inner edge of the disc, with perhaps a small contribution from a warm Comptonisation region (see e.g. \citealt{Done.2012}). This very steep soft X-ray spectrum is a feature of the newly discovered Quasi-Periodic Eruptions (QPE), seen in a very rare class of AGN. Intriguingly, the characteristic eruptions are clearly marked by a dramatic increase in soft X-ray excess emission (\citealt{Miniutti.2019, Arcodia.2021}).

\subsubsection{AGN Composite Spectra}
Finally, we compare the composite QSO spectrum from \citet{Francis.1991} with \src1, as shown in the cyan line in Figure~\ref{fig-optuv-compare} panels a6 and b6. In the optical band, the composite spectrum has stronger oxygen lines and weaker Fe {\sc ii} lines, indicating that the average Eddington ratio of the QSO sample is smaller. In the UV band, the composite spectrum has a continual shape similar to \src1\ and \phl1811, making it quite unlikely that the bend away from the standard disc shape in the UV is due to host galaxy reddening as this would require the \src1, \phl1811\ and the composite spectrum to have very similar E(B-V). Instead, the bend looks like an intrinsic spectral feature. The composite spectrum also has stronger and less blue-shifted emission lines than WLQs. For example, we measure Ly$\alpha$ + N {\sc v} REW = 49.7 \AA\ and C {\sc iv} REW = 32.6 \AA\ for the composite spectrum, which are even larger than \1h07\ and \rx04\ (Table~\ref{tab-comp-line}). The SED shape is very different to all the super-Eddington SEDs, with no strong evidence for a larger extreme-UV (EUV) component (Figure~\ref{fig-sed-compare}d).

\subsubsection{Weak-Line Seyfert Galaxy}
To summarize this section: \src1 has weak and blue-shifted UV emission lines sufficient to class it as a WLQ, but it has much lower black hole mass and higher mass accretion rate than most WLQs. While at the moment this is a unique object, we propose that such sources be called Weak Line Seyferts (WLS).

In comparison, objects with similarly super-Eddington mass accretion rates as \src1, but lower masses have UV emission lines which are not so weak and blue-shifted, so would not be defined as WLS, but rather classed as super-Eddington NLS1.

Most super-Eddington AGN with both low and high black hole masses show strong and complex soft and hard X-ray variability, plausibly due to an absorption/scattering in a clumpy wind. Our viewing angle with respect to this wind as well as the wind properties will determine the impact of this extrinsic variability (separate to the intrinsic variability of the corona/soft X-ray emission region) on the observed X-ray spectrum. This impacts on the derived optical-to-X-ray index $\alpha_{\rm ox}$, by reducing the observed X-ray flux when the wind material is in the line of sight, so that these events can be identified by the source being X-ray weaker than expected at a given 2500 \AA\ luminosity.

Interestingly, the dusty torus offers an independent viewing angle which is nearly edge-on (\citealt{Antonucci.1993}). It reprocesses the emission from the inner accretion flow and re-emits in the near infrared, thus the intensity of infrared torus emission may also provide clues about the properties of accretion flow. Indeed, Figure~\ref{fig-sed-compare}a shows that the relative infrared luminosities of \src1\ and \phl1811\ are very similar. More quantitatively, we adopt the optical-to-infrared index ($\alpha_{\rm optir}$, \citealt{Castello-Mor.2017}) as,
\begin{equation}
\alpha_{\rm optir} = - \frac{\rm{log}(L_{\rm 2500\textup{\AA}}/L_{\rm 5\mu m})}{\rm{log}(\nu_{\rm 2500\textup{\AA}}/\nu_{\rm 5\mu m})}
\end{equation}
where $L_{\rm 2500\textup{\AA}}$ and $L_{\rm 5\mu m}$ are the luminosities at 2500 \AA\ and 5 $\mu$m. Assuming an uncertainty of 10 per cent for the infrared luminosity, we find $\alpha_{\rm optir}$ is 0.67  $\pm$ 0.08 for \phl1811\ and 0.66 $\pm$ 0.08 for both SEDs of \src1\ (see Table~\ref{tab-comp-sed}). Hence it is likely that their torii also see similar SEDs from their separate accretion flows, which increases the global similarity between WLS and WLQs.

\begin{figure}
\centering
\includegraphics[trim=0.0in 0.4in 0.0in 0.0in, clip=1, scale=0.56]{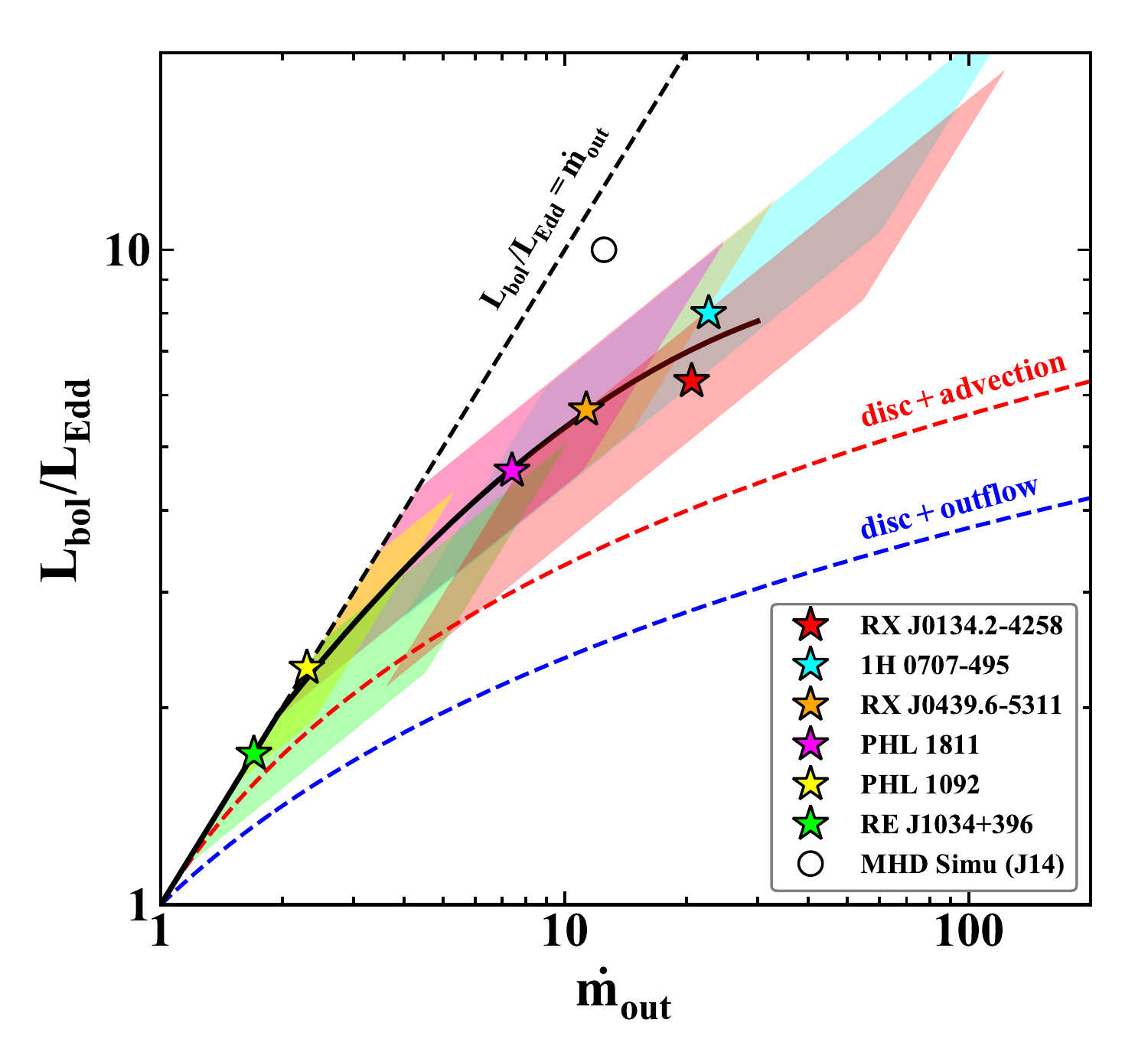}
\caption{The relation between the mass accretion rate ($\dot{m}_{\rm out}$) and Eddington ratio ($L_{\rm bol}/L_{\rm Edd}$) in the super-Eddington regime. Star symbols indicate the six super-Eddington AGN whose parameters are listed in Table~\ref{tab-comp-sed}. The circular symbol indicates the 3D MHD simulation of a super-Eddington accretion disc reported by \citet{Jiang.2014} (J14). The shaded regions indicate the uncertainty caused by the estimates of black hole mass and luminosity, assuming $L_{\rm bol}/L_{\rm Edd} \le \dot{m}_{\rm out}$. The red dash line indicates an accretion disc model with advection, as calculated by \citet{Poutanen.2007}, with the $x$ factor being 1.0 in the $L_{\rm bol}/L_{\rm Edd}=1+x\cdot$ln$(\dot{m}_{\rm out})$ equation. The blue dash line indicates an accretion disc model with outflow, with the $x$ factor being 0.6. The black dash line is the best-fit second-order polynomial relation for the six AGN (see Equation~\ref{eq-1}).}
\label{fig-mdot-eddr}
\end{figure}

\subsection{Efficiency of Super-Eddington Accretion Flows in AGN}
\label{sec-disc}
The global radiative efficiency ($\mu$) is a key parameter of the accretion flow around SMBHs, which indicates the fraction of accreted energy that is converted into radiation. $\mu$ can be estimated by measuring the difference between the observed Eddington ratio and mass accretion rate (\citealt{Davis.2011}). In Table~\ref{tab-comp-sed} we have estimated the radiative efficiency for \src1\ from its best-fit SEDs, which is found to be $\sim$ 30 per cent of the theoretical efficiency $\mu_{0}=0.057$ of a standard thin disc for $M_{\rm BH}=2\times10^{7}~M_{\odot}$ and zero spin. Likewise, we measure $\mu$ for the other AGN mentioned in this work based on their best-fit SEDs, which are shown in Table~\ref{tab-comp-sed}. 

We find that as $\dot{m}_{\rm out}$ increases, $\mu$ decreases significantly. This is qualitatively consistent with theoretical expectation that as the accretion flow becomes more super-Eddington, its properties deviate from the standard thin disc more significantly, the advection and wind may carry away a larger fraction of accretion energy, thus the global radiative efficiency becomes smaller. For instance, \citet{Poutanen.2007} derived a simple equation, $L_{\rm bol}/L_{\rm Edd}=1+x\cdot$ln$(\dot{m}_{\rm out})$, to describe the relation between the Eddington ratio and mass accretion rate for super-Eddington accretion flows. The $x$ factor is directly related to $\mu$. For a super-Eddington disc with advection but without outflow, the $x$ factor is 1.0; while for a disc with outflow but without advection, the $x$ factor is 0.6. In Figure~\ref{fig-mdot-eddr} we plot our small AGN sample on the parameter space of $\dot{m}_{\rm out}$ and $L_{\rm bol}/L_{\rm Edd}$, and compare them with various theoretical relations. We find that for a fixed $\dot{m}_{\rm out}$, the observed Eddington ratio is a factor of few higher than the disc+advection and disc+outflow model, and the deviation increases as AGN becomes more and more super-Eddington. This suggests that while the actual radiative efficiency of a super-Eddington accretion flow is significantly lower than that of a standard thin disc (\citealt{Shakura.1973}), it remains higher than those predicted by previous super-Eddington disc models.

Based on the six AGN in Figure~\ref{fig-mdot-eddr}, we perform second-order polynomial model fit to derive an empirical relation between $\dot{m}_{\rm out}$ and $L_{\rm bol}/L_{\rm Edd}$,
\begin{equation}
\label{eq-1}
\textup{log}(L_{\rm bol}/L_{\rm Edd})=a_0~[\textup{log}(\dot{m}_{\rm out})]^2+a_1~\textup{log}(\dot{m}_{\rm out})+a_2
\end{equation}
where $a_0 = -0.234^{+0.237}_{-0.136}$, $a_1 = 0.919^{+0.170}_{-0.189}$, and $a_2 = 0.044^{+0.009}_{-0.104}$ are derived from the parameter values and inferred uncertainties.
This equation can be used to infer the radiative efficiency for an AGN with $1.7\le \dot{m}_{\rm out} \lesssim 50$, but it is limited by the small sample size, and so it should be refined by future large sample studies.

Figure~\ref{fig-mdot-eddr} also shows the inferred uncertainty region for every source. The uncertainty of $\mu$ is mainly caused by the measurement accuracies of the black hole mass $M_{\rm BH}$, spin $a_{*}$ (we assumed spin 0 in this work as a conservative limit) and bolometric luminosity $L_{\rm bol}$. The typical measurement uncertainty of $L_{\rm bol}$ is a few tens of per cent for a well defined SED of an unobscured super-Eddington AGN with high-quality optical/UV and soft/hard X-ray data (e.g. \citealt{Jin.2012c, Jin.2013, Jin.2017b, Jin.2021}). The black hole spin is largely unknown, but \citet{Davis.2011} showed that spin can introduce a few tens of per cent uncertainty on $\dot{m}_{\rm out}$. However, this can be easily overwhelmed by the uncertainty of black hole mass estimate, because this uncertainty can be a factor of few, and we have $\dot{m}_{\rm out} \propto M_{\rm BH}^{-2}$ for an observed optical/UV luminosity, so the uncertainty propagated from $M_{\rm BH}$ to $\dot{m}_{\rm out}$ is square-amplified. The uncertainty regions in Figure~\ref{fig-mdot-eddr} are based on these black hole mass ranges and $\pm$ 50 per cent uncertainty of $L_{\rm bol}$, and we have also assumed that $L_{\rm bol}/L_{\rm Edd}$ cannot exceed $\dot{m}_{\rm out}$. We find that except \rej1034\ and \ph1092\ whose $\dot{m}_{\rm out}$ values are only 1.7 and 2.3, all the other uncertainty regions lie between the standard thin disc model and the disc+advection/outflow models. Therefore, considering various measurement uncertainties, it appears robust that the observed radiative efficiencies of our super-Eddington AGN sample are indeed lower than the prediction of standard thin disc model, but significantly higher than predictions of theoretical super-Eddington disc models.

It is also useful to compare our results with numerical simulations of super-Eddington accretion flows. For example, \citet{Jiang.2014} performed three-dimensional (3D) radiation magneto-hydrodynamical (MHD) simulations for an AGN with $M_{\rm BH}=4.2\times10^{6}~M_{\odot}$ and $\dot{m}_{\rm out}=12.5$. These parameters are comparable with those of \rx04. \citet{Jiang.2014} reported $L_{\rm bol}/L_{\rm Edd} \sim 10$, which then leads to $\mu = 0.045$, as shown by the star symbol in Figure~\ref{fig-mdot-eddr}. This radiative efficiency is significantly higher than those predicted by most theoretical models. It is caused by the inclusion of magnetic buoyancy in the simulation, which enhances the vertical advection of radiation, thereby increasing the global radiative efficiency. The efficiency we found for \rx04\ is 0.029, which is also much higher than previous theoretical models, but is $\sim$ 55 per cent lower than the 3D MHD simulation.

The observed correlation in Figure~\ref{fig-mdot-eddr} for the extreme super-Eddington regime is broadly consistent with the moderate correlation reported by \citet{Davis.2011} using a much larger sample of both sub and super-Eddington Palomar-Green quasars. \citet{Davis.2011} also reported a significant correlation between the radiative efficiency and the black hole mass, which can be explained by the mass-spin correlation predicted by the cosmic evolution of SMBH (\citealt{Fanidakis.2011}). We do not find such a relation in our small sample, which is probably because these six AGN only cover a narrow mass range, and so the effect of mass-spin correlation is negligible.

\subsection{A Proposed Picture for Super-Eddington Accretion Flows Depending on $M_{BH}$ and $\dot{m}_{\rm out}$}
\label{sec-disc2}
\begin{figure*}
\includegraphics[trim=1.6in 0in 1.5in 0in, clip=1, scale=0.65]{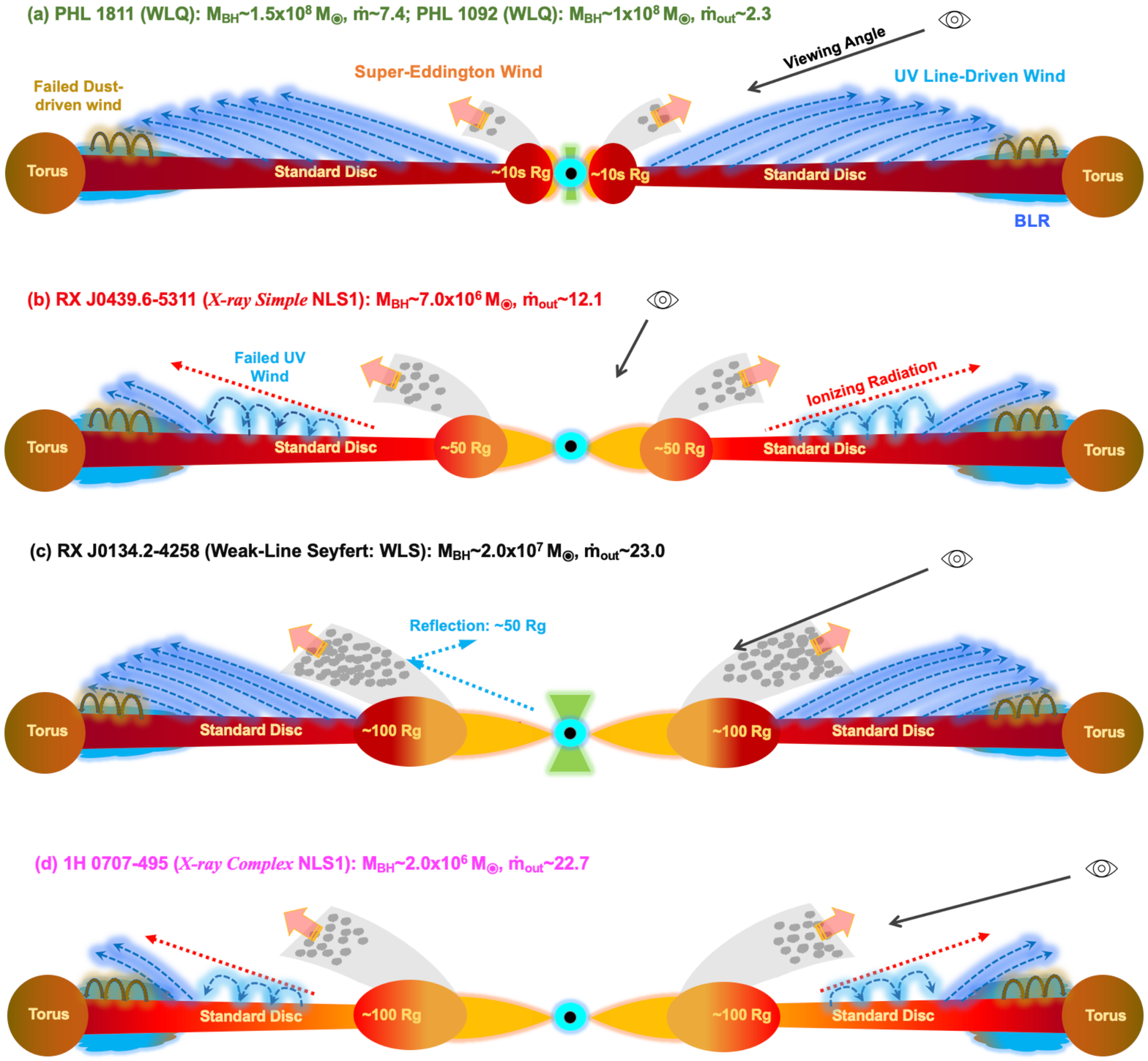}
\caption{The inferred structure of the super-Eddington accretion flows in the WLS \rxj0134\ and some super-Eddington NLS1s and WLQs discussed in this work. The disc structure depends on the black hole mass and mass accretion rate, which then leads to different properties of multi-wavelength emission. Another famous WLQ-like AGN in the local Universe is PSD 456, which has $M_{\rm BH}\sim 10^{9}M_{\odot}$ and $\dot{m}_{\rm out}\sim1$, and it may also belong to the scenario (a).}
\label{fig-discwind}
\end{figure*}

Super-Eddington NLS1s and WLQs are both characterized by their high mass accretion rates, but their black hole masses differ by more than one order of magnitude, thus comparing these two types of AGN can bring new insights on the super-Eddington accretion flow of SMBH. Based on their different multi-wavelength properties, a puffed-up inner disc scenario was proposed for both types of SMBH accretion systems (e.g. \citealt{Luo.2015} for WLQs and \citealt{Jin.2017b} for NLS1s). This was discussed even in the original paper of \citet{Shakura.1973}, where they show that the flow forms a quasi-spherical funnel
when the disc reaches the local Eddington limit, within the spherization radius $r_{\rm sp}~=~R_{\rm sp}/R_{\rm in}~\sim~\dot{m}_{\rm out}$ where $R_{\rm in}$ is the inner radius (see also \citealt{Poutanen.2007}, Kubota \& Done 2019). We can derive a rough estimate of this radius for each object.
Such an inner disc structure will produce a geometric collimation of the inner disc/soft X-ray/hard X-ray emission. This inclination dependence will be enhanced by any wind from the super-Eddington flow, and strong evidence for a wind is seen in the rapid and complex X-ray variability. However, both the funnel and the super-Eddington wind 
might be expected to depend only on $\dot{m}_{\rm out}$ rather than on mass,  yet we showed examples in the previous section where 
sources at similar high Eddington fractions have weaker/more blue-shifted UV lines at higher masses, so there should be another intrinsic factor
at work as well. 

One possibility is that there is an additional wind from the disc due to UV line driving. This is already implicated in the WLQ by the fact that the UV lines (C {\sc iv} and Ly$\alpha$+N {\sc v}) are blue-shifted. UV line driving is sensitively dependent on the disc temperature which depends on $M_{\rm BH}$ as well as $\dot{m}_{\rm out}$. 
A continuum SED which peaks in the UV (as in the mildly super-Eddington high mass AGN) is much more efficient in UV line driving than one that peaks in the EUV/soft X-rays
(as in the strongly super-Eddington low-mass NLS1s). Hence this gives a component which depends on mass as well as $\dot{m}_{\rm out}$. Based on the above ideas, we compare the inferred structure of super-Eddington accretion discs for different black hole masses and mass accretion rates, which is shown in Figure~\ref{fig-discwind}.

\begin{figure}
\centering
\includegraphics[trim=0.0in 0.2in 0.0in 0.0in, clip=1, scale=0.48]{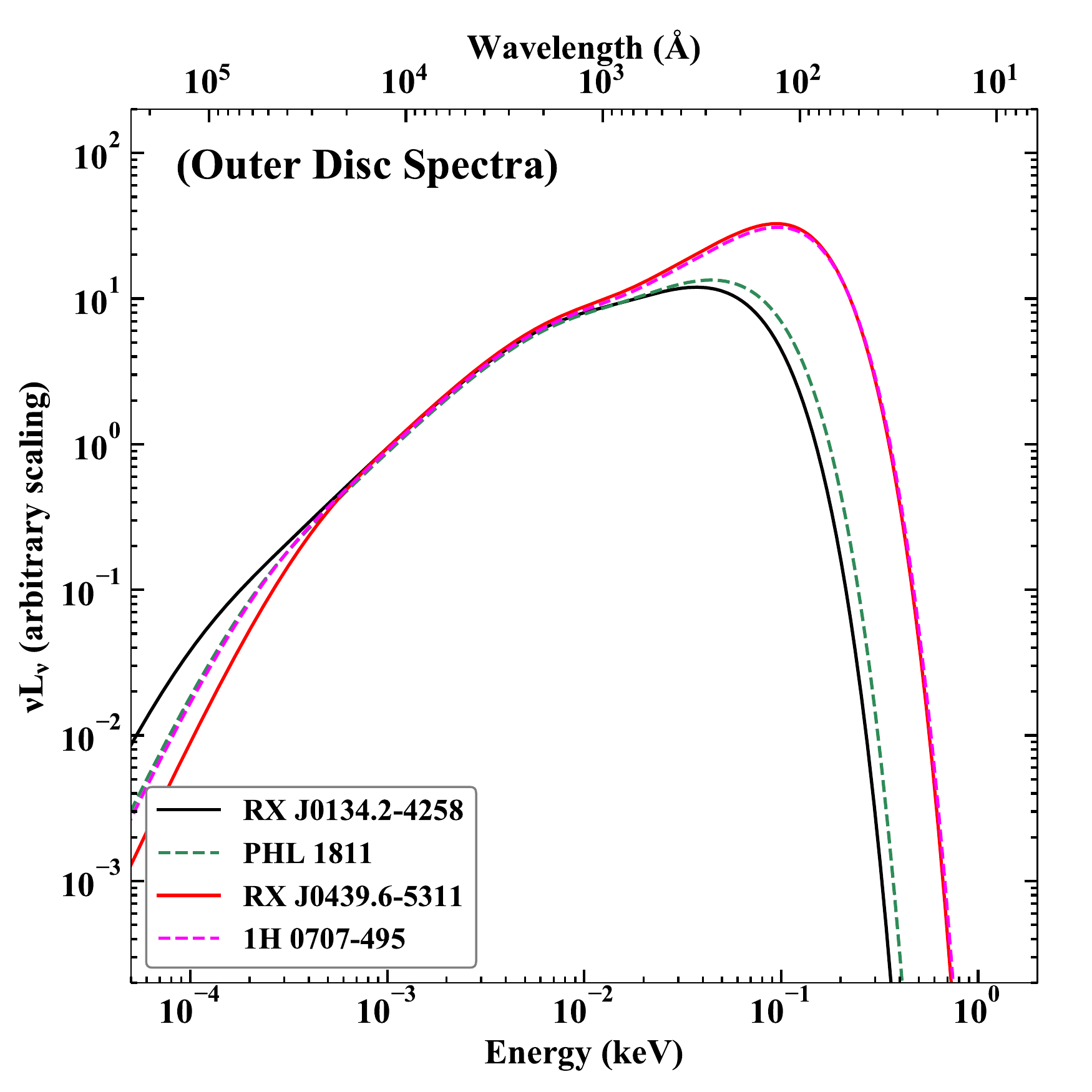}
\caption{Comparison of the inferred SEDs produced by the disc region outside the spherization radius ($R_{\rm sp} \sim R_{\rm in}~\dot{m}_{\rm out}=6 R_{\rm g}~\dot{m}_{\rm out}$) of the four super-Eddington AGN. The {\tt optxagnf} model is adopted and is truncated at $R_{\rm sp}$. The input $M_{\rm BH}$ and $\dot{m}_{\rm out}$ are from Table~\ref{tab-comp-sed}. Since only the shape of SED is concerned, arbitrary scaling factors have been applied to make the four SEDs matching in the optical. It is clear that the SED shapes of the outer discs of \src1\ and \phl1811\ are very similar to each other, which explains their similar weak-line properties; while the SED shapes of \rx04\ and \1h07\ are also similar to each other and contain a lot more high-energy flux, which makes the UV line-driven wind inefficient in these two sources.}
\label{fig-outer-disc}
\end{figure}

The first row is the puffed-up disc scenario proposed for WLQs such as \phl1811\ and \ph1092\ (e.g. \citealt{Wu.2012, Luo.2015, Ni.2018}). Since their mass accretion rates are close to or slightly above the Eddington limit, the puffed up region is quite small (a few tens of $R_{\rm g}$), and the Eddington wind is not very powerful. However, these structures shield the rest of the disc from the hottest emission (inner funnel and X-ray corona). The maximum outer disc temperature is given at the radius where the puffed up region starts. To infer the SED shape outside the spherization radius, we adopt the {\tt optxagnf} model (\citealt{Done.2012}) and take the best-fit $M_{\rm BH}$ and $\dot{m}_{\rm out}$ listed in Table~\ref{tab-comp-sed} as inputs\footnote{As a rough comparison, we take the average $\dot{m}_{\rm out}$ of 25 for \1h07, and 20.6 for \src1.}, and then truncate the SED at $R_{\rm sp}$. As shown in Figure~\ref{fig-outer-disc}, for WLQs with masses of $\sim 10^8$, this outer disc emits in the UV, and is just below Eddington. This should power an extremely strong UV line-driven disc wind. 
This shields the standard BLR from the outer disc from the inner UV/X-ray emission, so the standard BLR which makes the core of the lines is weak, so the UV lines are dominated by the wind emission, giving the characteristic blueshift of the BLR line profile. In Figure~\ref{fig-discwind}, we also show the failed dust-driven wind from the model of \citet{Czerny.2011}, where it is suggested that the BLR is itself a (failed) wind, but driven by dust rather than by the UV\footnote{This disc scenario may also be applicable for another famous AGN called PDS 456. It is a low-redshift AGN located at $z=0.184$ with $M_{\rm BH}\sim\times10^{9}~M_{\odot}$ and $L_{\rm bol}/L_{\rm Edd}\sim 1.0$ (\citealt{Simpson.1999, Reeves.2000}). It is famous for showing extreme X-ray variability and X-ray absorption features indicating ultra-fast outflows (\citealt{Reeves.2020} and references therein). Besides, it is known to show weak [O {\sc iii}] $\lambda 5007$ line with REW $<$ 2 \AA\ (\citealt{Simpson.1999}), as well as weak UV high-ionization lines such as C {\sc iv} REW $=14.7$ \AA\ and broad UV absorption lines (\citealt{Hamann.2018}). These multi-wavelength properties are similar to \src1\ and \phl1811, although its C {\sc iv} is still stronger than the definition of WLQ. PDS 456 also shows a low level of radio emission, although it can be classified as a radio-quiet AGN (\citealt{Vignali.2000, Yang.2021}).}.

The second row of Figure~\ref{fig-discwind} shows the disc scenario for the archetypal super-Eddington NLS1 \rx04\ (\citealt{Jin.2017b}). Comparing to \phl1811, its black hole mass is one order of magnitude lower, and its mass accretion rate is one order of magnitude higher, thus we expect the disc to be much hotter ($T_{\rm eff}^{4} \propto M_{\rm BH}^{-1}~\dot{m}$)
and the puffed-up disc region to be larger ($\sim$ 50 $R_{\rm g}$), and so the super-Eddington disc wind may be more powerful. This again shields the outer disc from the hottest parts of the disc, but the disc temperature just outside of the funnel region is 
somewhat hotter than seen in WLQs (see Figure~\ref{fig-outer-disc}). This stronger emission below 200\AA\ could be enough to over-ionise the disk wind, so the UV line driving is not efficient. The UV wind then does not shield the core of BLR as efficiently, so the core of the UV lines (e.g. C {\sc iv}) are stronger, as well as the blue-wing of the lines being weaker (due to the weaker UV wind), so these line profiles do not meet the criteria for a WLQ.

The above picture is further supported by the multi-wavelength properties of the WLS \src1, whose black hole mass and mass accretion rates are both higher than \rx04, as shown in the third row of Figure~\ref{fig-discwind}. In this case, higher mass accretion rate means that the disc can begin to puff up at an even larger radius of $\sim$ 100 $R_{\rm g}$, and so an even stronger super-Eddington wind is expected. This clumpy wind absorbs the soft X-rays of \src1\ and causes its drastic X-ray variability, similar to those observed in \ph1092\ and PDS 456.
But the emission from the disk outside of the funnel is now similar to that of the WLQ (see Figure~\ref{fig-outer-disc}), so this can have a similarly strong UV line driven wind, and so weak core lines from the BLR, and UV lines dominated by the UV line driven disc wind. 
Then the UV line-driving mechanism still works efficiently at outer/cooler disc region, making \src1\ the so far unique WLS. This also explains why \src1\ and \phl1811\ show similar infrared luminosity from hot torus relative to their optical luminosity.

A jet is also drawn in this picture, because the radio-loudness of \rej1034\ was reported to be $\sim~17$, and so it was a marginally radio-loud NLS1 more than one decade ago (\citealt{Gelbord.2009}). But our latest observation campaign shows that now it becomes a radio-quiet source, and so its jet emission may be episodic (see more details in Paper-III).

The last row of Figure~\ref{fig-discwind} shows the disc scenario for \1h07, whose $\dot{m}_{\rm out}$ is similar to \src1, but with one order-of-magnitude lower black hole mass. In this case, the accretion disc is the hottest among all the AGN mentioned above, and so the super-Eddington wind is also the strongest, which leads to drastic X-ray variability (e.g. \citealt{Hagino.2016, Done.2016, Boller.2021, Parker.2021}). The disc emission outside of the funnel region is now similar to \rx04\ (see Figure~\ref{fig-outer-disc}), just a little too high for the UV line driving to be efficient, so the blue-shifted line from the wind is not strong and the core BLR is not shielded. This explains why \1h07\ does not turn out to be a WLS.

Finally, we emphasize that our study of the similarities and differences between super-Eddington NLS1s and WLQs is just at the beginning. More similarities/differences may be found by future studies, including detailed photoionization analyses of the BLRs of super-Eddington NLS1s and WLQs assuming different illuminating SEDs. If the above scenarios are generally correct, we speculate that further studies may find more WLS, as well as more WLQs showing drastic X-ray variability.

\section{Summary and Conclusions}
\label{sec-conclusion}
We carried out a multi-wavelength campaign on the enigmatic super-Eddington NLS1 \rxj0134\ from radio to optical/UV to X-rays, using both space and ground-based telescopes. In this work, we present a detailed optical/UV spectroscopic analysis as well as broadband SED analysis from infrared to X-rays, and compare these multi-wavelength properties with
other super-Eddington NLS1s and WLQs,
thereby yielding deeper understanding about the super-Eddington accretion flows around SMBHs. The main results of this paper are summarized below.

\begin{itemize}
\item the optical/UV spectra of \rxj0134\ show extremely weak UV high-ionization lines such as C {\sc iv}, Si {\sc iv} and N {\sc v}, which are consistent with the definition of a WLQ. Together with other similarities such as the drastic X-ray variability and optical-to-infrared flux ratio, we propose \rxj0134\ as a new category of AGN, namely the weak-line Seyfert (WLS), which can be considered as the low mass and higher mass accretion rate counterpart of WLQs.

\item we build the broadband SED of \rxj0134, which shows that the soft excess of \rxj0134\ is more than one order of magnitude weaker than in X-ray {\it simple} super-Eddington NLS1s. For a preferred black hole mass of $2 \times10^{7}M_{\sun}$ and $a_{*}=0$, the mass accretion rate $\dot{m}_{\rm out}$ is found to be $\sim$ 20, $L_{\rm bol}/L_{\rm Edd}$ is $\sim$ 6,
so the radiative efficiency is only $\sim$ 30 per cent of that of a standard thin disc. 

\item by performing a systematic comparison within a small but representative super-Eddington AGN sample, we find that the most extreme NLS1s with similarly large $\dot{m}_{\rm out}$ but smaller masses than \rxj0134\ do not show similarly weak and wind-dominated UV high-ionization lines as WLQs and WLS do. Thus the properties of accretion flow should depend on both black hole mass and mass accretion rate.

\item in the super-Eddington regime, the observed global radiative efficiency of the accretion flow decreases significantly as the mass accretion rate increases. The measured efficiency is higher than expected from the standard thin disc model and previous super-Eddington disc models, but lower than previous 3D MHD simulations of super-Eddington AGN discs.

\item we propose a picture to show the dependence of super-Eddington accretion flows on the black hole mass, mass accretion rate and inclination angle, which can be used to qualitatively understand the multi-wavelength spectral differences between different subtypes of super-Eddington AGN, including super-Eddington NLS1s, WLQs and WLS.
\end{itemize}

The multi-wavelength long-term variability of \rxj0134\ from radio to X-rays as revealed by our {\it ATCA} observations and ongoing \swift\ observations will be presented in our next paper (Paper-III).

\section*{Acknowledgements}
CJ thanks Niel Brandt, Jianfeng Wu, Jianmin Wang, Yanrong Li and Erlin Qiao for valuable discussions, and thanks Peng Jiang, Hongyan Zhou and Weimin Yuan for helping with the coordination of optical observations. We thank Murilo Marinello for sharing with us the high-quality optical/UV/IR spectra of \ph1092. We thank the teams of the {\it Swift} satellite and the Siding Spring Observatory for approving and conducting the target-of-opportunity observations, as well as helping with the data reduction. We also thank the \xmm\ team for helping to investigate instrumental issues related to the OM data.

CJ acknowledges the National Natural Science Foundation of China through grant 11873054, and the support by the Strategic Pioneer Program on Space Science, Chinese Academy of Sciences through grant XDA15052100. CD acknowledges the Science and Technology Facilities Council (STFC) through grant ST/T000244/1 for support. HL acknowledges the support by Chinese Postdoctoral Science Foundation (2021M693203), and the National Natural Science Foundation of China through grant 12103061.

This work is based on observations conducted by \xmm, an ESA
science mission with instruments and contributions directly funded by
ESA Member States and the USA (NASA). This work also makes use of data from the \nustar\ mission, a project led by the California Institute of Technology, managed by the Jet Propulsion Laboratory, and funded by the National Aeronautics and Space Administration. This research has made use of the XRT Data Analysis Software (XRTDAS) developed under the responsibility of the ASI Science Data Center (ASDC), Italy.


\section*{Data Availability}
The data underlying this article are publicly available from the High Energy Astrophysics Science Archive Research Center (HEASARC) at https://heasarc.gsfc.nasa.gov, the \xmm\ Science Archive (XSA) at https://www.cosmos.esa.int/web/xmm-newton/xsa, the Barbara A. Mikulski Archive for Space Telescopes (MAST) at https://mast.stsci.edu/portal/Mashup/Clients/Mast/Portal.html, the Sloan Digital Sky Survey (SDSS) at http://skyserver.sdss.org/dr12/en/home.aspx. The SSO optical spectrum in this article will be shared on reasonable request to the corresponding author.




\begin{thebibliography}{99}

\bibitem[\protect\citeauthoryear{Abramowicz et al.}{1988}]{Abramowicz.1988} Abramowicz M.~A., Czerny B., Lasota J.~P., Szuszkiewicz E., 1988, ApJ, 332, 646

\bibitem[\protect\citeauthoryear{Alston et al.}{2014}]{Alston.2014} Alston W.~N., Markeviciute J., Kara E., Fabian A.~C., Middleton M., 2014, MNRAS, 445, L16

\bibitem[\protect\citeauthoryear{Antonucci}{1993}]{Antonucci.1993} Antonucci R., 1993, ARA\&A, 31, 473

\bibitem[\protect\citeauthoryear{Arcodia et al.}{2021}]{Arcodia.2021} Arcodia R., Merloni A., Nandra K., Buchner J., Salvato M., Pasham D., Remillard R., et al., 2021, Natur, 592, 704

\bibitem[\protect\citeauthoryear{Arnaud}{1996}]{Arnaud.1996} Arnaud K.~A., 1996, ASPC, 101, 17

\bibitem[\protect\citeauthoryear{Baskin \& Laor}{2018}]{Baskin.2018} Baskin A., Laor A., 2018, MNRAS, 474, 1970

\bibitem[\protect\citeauthoryear{Blackburn}{1995}]{Blackburn.1995} Blackburn J.~K., 1995, ASPC, 77, 367

\bibitem[\protect\citeauthoryear{Boller, Brandt \& Fink}{1996}]{Boller.1996} Boller T., Brandt W.~N., Fink H., 1996, A\&A, 305, 53

\bibitem[\protect\citeauthoryear{Boller et al.}{2021}]{Boller.2021} Boller T., Liu T., Weber P., Arcodia R., Dauser T., Wilms J., Nandra K., et al., 2021, A\&A, 647, A6

\bibitem[\protect\citeauthoryear{Boroson}{2002}]{Boroson.2002} Boroson T.~A., 2002, ApJ, 565, 78

\bibitem[\protect\citeauthoryear{Bottorff et al.}{1997}]{Bottorff.1997} Bottorff M., Korista K.~T., Shlosman I., Blandford R.~D., 1997, ApJ, 479, 200

\bibitem[\protect\citeauthoryear{Brandt, Mathur \& Elvis}{1997}]{Brandt.1997} Brandt W.~N., Mathur S., Elvis M., 1997, MNRAS, 285, L25

\bibitem[\protect\citeauthoryear{Castell{\'o}-Mor et al.}{2017}]{Castello-Mor.2017} Castell{\'o}-Mor N., Kaspi S., Netzer H., Du P., Hu C., Ho L.~C., Bai J.-M., et al., 2017, MNRAS, 467, 1209

\bibitem[\protect\citeauthoryear{Coatman et al.}{2016}]{Coatman.2016} Coatman L., Hewett P.~C., Banerji M., Richards G.~T., 2016, MNRAS, 461, 647

\bibitem[\protect\citeauthoryear{Collinson et al.}{2015}]{Collinson.2015} Collinson J.~S., Ward M.~J., Done C., Landt H., Elvis M., McDowell J.~C., 2015, MNRAS, 449, 2174

\bibitem[\protect\citeauthoryear{Collinson et al.}{2017}]{Collinson.2017} Collinson J.~S., Ward M.~J., Landt H., Done C., Elvis M., McDowell J.~C., 2017, MNRAS, 465, 358

\bibitem[\protect\citeauthoryear{Crummy et al.}{2006}]{Crummy.2006} Crummy J., Fabian A.~C., Gallo L., Ross R.~R., 2006, MNRAS, 365, 1067

\bibitem[\protect\citeauthoryear{Czerny \& Hryniewicz}{2011}]{Czerny.2011} Czerny B., Hryniewicz K., 2011, A\&A, 525, L8

\bibitem[\protect\citeauthoryear{Czerny et al.}{2015}]{Czerny.2015} Czerny B., Modzelewska J., Petrogalli F., Pych W., Adhikari T.~P., {\.Z}ycki P.~T., Hryniewicz K., et al., 2015, AdSpR, 55, 1806

\bibitem[\protect\citeauthoryear{Czerny et al.}{2016}]{Czerny.2016} Czerny B., You B., Kurcz A., {\'S}redzi{\'n}ska J., Hryniewicz K., Niko{\l}ajuk M., Krupa M., et al., 2016, A\&A, 594, A102

\bibitem[\protect\citeauthoryear{Davis \& Laor}{2011}]{Davis.2011} Davis S.~W., Laor A., 2011, ApJ, 728, 98

\bibitem[\protect\citeauthoryear{Diamond-Stanic et al.}{2009}]{Diamond-Stanic.2009} Diamond-Stanic A.~M., Fan X., Brandt W.~N., Shemmer O., Strauss M.~A., Anderson S.~F., Carilli C.~L., et al., 2009, ApJ, 699, 782

\bibitem[\protect\citeauthoryear{Done et al.}{2012}]{Done.2012} Done C., Davis S.~W., Jin C., Blaes O., Ward M., 2012, MNRAS, 420, 1848

\bibitem[\protect\citeauthoryear{Done et al.}{2013}]{Done.2013} Done C., Jin C., Middleton M., Ward M., 2013, MNRAS, 434, 1955

\bibitem[\protect\citeauthoryear{Done \& Jin}{2016}]{Done.2016} Done C., Jin C., 2016, MNRAS, 460, 1716

\bibitem[\protect\citeauthoryear{Dong et al.}{2008}]{Dong.2008} Dong X., Wang T., Wang J., Yuan W., Zhou H., Dai H., Zhang K., 2008, MNRAS, 383, 581

\bibitem[\protect\citeauthoryear{Du et al.}{2018}]{Du.2018} Du P., Zhang Z.-X., Wang K., Huang Y.-K., Zhang Y., Lu K.-X., Hu C., et al., 2018, ApJ, 856, 6

\bibitem[\protect\citeauthoryear{Du \& Wang}{2019}]{Du.2019} Du P., Wang J.-M., 2019, ApJ, 886, 42

\bibitem[\protect\citeauthoryear{Edelson et al.}{2015}]{Edelson.2015} Edelson R., Gelbord J.~M., Horne K., McHardy I.~M., Peterson B.~M., Ar{\'e}valo P., Breeveld A.~A., et al., 2015, ApJ, 806, 12

\bibitem[\protect\citeauthoryear{Elvis et al.}{1994}]{Elvis.1994} Elvis M., Wilkes B.~J., McDowell J.~C., Green R.~F., Bechtold J., Willner S.~P., Oey M.~S., et al., 1994, ApJS, 95, 1

\bibitem[\protect\citeauthoryear{Fabian et al.}{2013}]{Fabian.2013} Fabian A.~C., Kara E., Walton D.~J., Wilkins D.~R., Ross R.~R., Lozanov K., Uttley P., et al., 2013, MNRAS, 429, 2917

\bibitem[\protect\citeauthoryear{Fan et al.}{1999}]{Fan.1999} Fan X., Strauss M.~A., Gunn J.~E., Lupton R.~H., Carilli C.~L., Rupen M.~P., Schmidt G.~D., et al., 1999, ApJL, 526, L57

\bibitem[\protect\citeauthoryear{Fanidakis et al.}{2011}]{Fanidakis.2011} Fanidakis N., Baugh C.~M., Benson A.~J., Bower R.~G., Cole S., Done C., Frenk C.~S., 2011, MNRAS, 410, 53

\bibitem[\protect\citeauthoryear{Fitzpatrick \& Massa}{2007}]{Fitzpatrick.2007} Fitzpatrick E.~L., Massa D., 2007, ApJ, 663, 320

\bibitem[\protect\citeauthoryear{Francis et al.}{1991}]{Francis.1991} Francis P.~J., Hewett P.~C., Foltz C.~B., Chaffee F.~H., Weymann R.~J., Morris S.~L., 1991, ApJ, 373, 465

\bibitem[\protect\citeauthoryear{Fuller et al.}{2016}]{Fuller.2016} Fuller L., Lopez-Rodriguez E., Packham C., Ramos-Almeida C., Alonso-Herrero A., Levenson N.~A., Radomski J., et al., 2016, MNRAS, 462, 2618

\bibitem[\protect\citeauthoryear{Gallo}{2006}]{Gallo.2006} Gallo L.~C., 2006, MNRAS, 368, 479

\bibitem[\protect\citeauthoryear{Gehrels et al.}{2004}]{Gehrels.2004} Gehrels N., Chincarini G., Giommi P., Mason K.~O., Nousek J.~A., Wells A.~A., White N.~E., et al., 2004, ApJ, 611, 1005

\bibitem[\protect\citeauthoryear{Gelbord, Mullaney \& Ward}{2009}]{Gelbord.2009} Gelbord J.~M., Mullaney J.~R., Ward M.~J., 2009, MNRAS, 397, 172

\bibitem[\protect\citeauthoryear{Gierli{\'n}ski et al.}{2008}]{Gierlinski.2008} Gierli{\'n}ski M., Middleton M., Ward M., Done C., 2008, Natur, 455, 369

\bibitem[\protect\citeauthoryear{Goad, Korista \& Ruff}{2012}]{Goad.2012} Goad M.~R., Korista K.~T., Ruff A.~J., 2012, MNRAS, 426, 3086

\bibitem[\protect\citeauthoryear{Grier et al.}{2017}]{Grier.2017} Grier C.~J., Pancoast A., Barth A.~J., Fausnaugh M.~M., Brewer B.~J., Treu T., Peterson B.~M., 2017, ApJ, 849, 146

\bibitem[\protect\citeauthoryear{Grupe et al.}{2000}]{Grupe.2000} Grupe D., Leighly K.~M., Thomas H.-C., Laurent-Muehleisen S.~A., 2000, A\&A, 356, 11

\bibitem[\protect\citeauthoryear{Grupe et al.}{2010}]{Grupe.2010} Grupe D., Komossa S., Leighly K.~M., Page K.~L., 2010, ApJS, 187, 64

\bibitem[\protect\citeauthoryear{Hagino et al.}{2016}]{Hagino.2016} Hagino K., Odaka H., Done C., Tomaru R., Watanabe S., Takahashi T., 2016, MNRAS, 461, 3954

\bibitem[\protect\citeauthoryear{Hamann et al.}{2018}]{Hamann.2018} Hamann F., Chartas G., Reeves J., Nardini E., 2018, MNRAS, 476, 943

\bibitem[\protect\citeauthoryear{Harrison et al.}{2013}]{Harrison.2013} Harrison F.~A., Craig W.~W., Christensen F.~E., Hailey C.~J., Zhang W.~W., Boggs S.~E., Stern D., et al., 2013, ApJ, 770, 103

\bibitem[\protect\citeauthoryear{Ho \& Kim}{2014}]{Ho.2014} Ho L.~C., Kim M., 2014, ApJ, 789, 17

\bibitem[\protect\citeauthoryear{Jansen et al.}{2001}]{Jansen.2001} Jansen F., Lumb D., Altieri B., Clavel J., Ehle M., Erd C., Gabriel C., et al., 2001, A\&A, 365, L1

\bibitem[\protect\citeauthoryear{Jiang, Stone \& Davis}{2014}]{Jiang.2014} Jiang Y.-F., Stone J.~M., Davis S.~W., 2014, ApJ, 796, 106

\bibitem[\protect\citeauthoryear{Jiang, Davis \& Stone}{2016}]{Jiang.2016} Jiang Y.-F., Davis S.~W., Stone J.~M., 2016, ApJ, 827, 10

\bibitem[\protect\citeauthoryear{Jin et al.}{2012a}]{Jin.2012a} Jin C., Ward M., Done C., Gelbord J., 2012, MNRAS, 420, 1825

\bibitem[\protect\citeauthoryear{Jin, Ward \& Done}{2012b}]{Jin.2012b} Jin C., Ward M., Done C., 2012, MNRAS, 422, 3268

\bibitem[\protect\citeauthoryear{Jin, Ward \& Done}{2012c}]{Jin.2012c} Jin C., Ward M., Done C., 2012, MNRAS, 425, 907

\bibitem[\protect\citeauthoryear{Jin et al.}{2013}]{Jin.2013} Jin C., Done C., Middleton M., Ward M., 2013, MNRAS, 436, 3173

\bibitem[\protect\citeauthoryear{Jin, Done \& Ward}{2016}]{Jin.2016} Jin C., Done C., Ward M., 2016, MNRAS, 455, 691

\bibitem[\protect\citeauthoryear{Jin, Done \& Ward}{2017a}]{Jin.2017a} Jin C., Done C., Ward M., 2017, MNRAS, 468, 3663

\bibitem[\protect\citeauthoryear{Jin et al.}{2017b}]{Jin.2017b} Jin C., Done C., Ward M., Gardner E., 2017, MNRAS, 471, 706

\bibitem[\protect\citeauthoryear{Jin, Done \& Ward}{2020}]{Jin.2020} Jin C., Done C., Ward M., 2020, MNRAS, 495, 3538

\bibitem[\protect\citeauthoryear{Jin, Done \& Ward}{2021}]{Jin.2021} Jin C., Done C., Ward M., 2021, MNRAS, 500, 2475

\bibitem[\protect\citeauthoryear{Jin et al.}{2022}]{Jin.2022} Jin C., Done C., Ward M., Panessa F., Liu B., Liu H., 2022, MNRAS, 512, 5642

\bibitem[\protect\citeauthoryear{Kaspi et al.}{2000}]{Kaspi.2000} Kaspi S., Smith P.~S., Netzer H., Maoz D., Jannuzi B.~T., Giveon U., 2000, ApJ, 533, 631

\bibitem[\protect\citeauthoryear{Kubota \& Done}{2018}]{Kubota.2018} Kubota A., Done C., 2018, MNRAS, 480, 1247

\bibitem[\protect\citeauthoryear{Kubota \& Done}{2019}]{Kubota.2019} Kubota A., Done C., 2019, MNRAS, 489, 524

\bibitem[\protect\citeauthoryear{Kwan \& Krolik}{1979}]{Kwan.1979} Kwan J., Krolik J.~H., 1979, ApJL, 233, L91

\bibitem[\protect\citeauthoryear{Kynoch et al.}{2018}]{Kynoch.2018} Kynoch D., Landt H., Ward M.~J., Done C., Gardner E., Boisson C., Arrieta-Lobo M., et al., 2018, MNRAS, 475, 404

\bibitem[\protect\citeauthoryear{Landt et al.}{2019}]{Landt.2019} Landt H., Ward M.~J., Kynoch D., Packham C., Ferland G.~J., Lawrence A., Pott J.-U., et al., 2019, MNRAS, 489, 1572

\bibitem[\protect\citeauthoryear{Laor et al.}{1997}]{Laor.1997} Laor A., Fiore F., Elvis M., Wilkes B.~J., McDowell J.~C., 1997, ApJ, 477, 93

\bibitem[\protect\citeauthoryear{Leighly et al.}{2007a}]{Leighly.2007a} Leighly K.~M., Halpern J.~P., Jenkins E.~B., Grupe D., Choi J., Prescott K.~B., 2007, ApJ, 663, 103

\bibitem[\protect\citeauthoryear{Leighly et al.}{2007b}]{Leighly.2007b} Leighly K.~M., Halpern J.~P., Jenkins E.~B., Casebeer D., 2007, ApJS, 173, 1

\bibitem[\protect\citeauthoryear{Li et al.}{2018}]{Li.2018} Li Y.-R., Songsheng Y.-Y., Qiu J., Hu C., Du P., Lu K.-X., Huang Y.-K., et al., 2018, ApJ, 869, 137

\bibitem[\protect\citeauthoryear{Liu et al.}{2019}]{Liu.2019} Liu H.-Y., Liu W.-J., Dong X.-B., Zhou H., Wang T., Lu H., Yuan W., 2019, ApJS, 243, 21

\bibitem[\protect\citeauthoryear{Londish et al.}{2004}]{Londish.2004} Londish D., Heidt J., Boyle B.~J., Croom S.~M., Kedziora-Chudczer L., 2004, MNRAS, 352, 903

\bibitem[\protect\citeauthoryear{Lu \& Yu}{2001}]{Lu.2001} Lu Y., Yu Q., 2001, MNRAS, 324, 653

\bibitem[\protect\citeauthoryear{Luo et al.}{2015}]{Luo.2015} Luo B., Brandt W.~N., Hall P.~B., Wu J., Anderson S.~F., Garmire G.~P., Gibson R.~R., et al., 2015, ApJ, 805, 122

\bibitem[\protect\citeauthoryear{Lusso et al.}{2010}]{Lusso.2010} Lusso E., Comastri A., Vignali C., Zamorani G., Brusa M., Gilli R., Iwasawa K., et al., 2010, A\&A, 512, A34

\bibitem[\protect\citeauthoryear{Lusso \& Risaliti}{2016}]{Lusso.2016} Lusso E., Risaliti G., 2016, ApJ, 819, 154

\bibitem[\protect\citeauthoryear{Magdziarz et al.}{1998}]{Magdziarz.1998} Magdziarz P., Blaes O.~M., Zdziarski A.~A., Johnson W.~N., Smith D.~A., 1998, MNRAS, 301, 179

\bibitem[\protect\citeauthoryear{Marinello et al.}{2020}]{Marinello.2020} Marinello M., Rodr{\'\i}guez-Ardila A., Marziani P., Sigut A., Pradhan A., 2020, MNRAS, 494, 4187

\bibitem[\protect\citeauthoryear{Mart{\'\i}nez-Paredes et al.}{2017}]{Martinez.2017} Mart{\'\i}nez-Paredes M., Aretxaga I., Alonso-Herrero A., Gonz{\'a}lez-Mart{\'\i}n O., Lop{\'e}z-Rodr{\'\i}guez E., Ramos Almeida C., Asensio Ramos A., et al., 2017, MNRAS, 468, 2

\bibitem[\protect\citeauthoryear{Mathews, Blumenthal \& Grandi}{1980}]{Mathews.1980} Mathews W.~G., Blumenthal G.~R., Grandi S.~A., 1980, ApJ, 235, 971

\bibitem[\protect\citeauthoryear{Mathur, Kuraszkiewicz \& Czerny}{2001}]{Mathur.2001} Mathur S., Kuraszkiewicz J., Czerny B., 2001, NewA, 6, 321

\bibitem[\protect\citeauthoryear{Matthews et al.}{2020}]{Matthews.2020} Matthews J.~H., Knigge C., Higginbottom N., Long K.~S., Sim S.~A., Mangham S.~W., Parkinson E.~J., et al., 2020, MNRAS, 492, 5540

\bibitem[\protect\citeauthoryear{McDowell et al.}{1995}]{McDowell.1995} McDowell J.~C., Canizares C., Elvis M., Lawrence A., Markoff S., Mathur S., Wilkes B.~J., 1995, ApJ, 450, 585

\bibitem[\protect\citeauthoryear{McKee}{1995}]{McKee.1995} McKee C.~F., 1995, ASPC, 80, 292

\bibitem[\protect\citeauthoryear{Middleton, Uttley \& Done}{2011}]{Middleton.2011} Middleton M., Uttley P., Done C., 2011, MNRAS, 417, 250

\bibitem[\protect\citeauthoryear{Miller et al.}{2007}]{Miller.2007} Miller L., Turner T.~J., Reeves J.~N., George I.~M., Kraemer S.~B., Wingert B., 2007, A\&A, 463, 131

\bibitem[\protect\citeauthoryear{Miniutti \& Fabian}{2004}]{Miniutti.2004} Miniutti G., Fabian A.~C., 2004, MNRAS, 349, 1435

\bibitem[\protect\citeauthoryear{Miniutti et al.}{2012}]{Miniutti.2012} Miniutti G., Brandt W.~N., Schneider D.~P., Fabian A.~C., Gallo L.~C., Boller T., 2012, MNRAS, 425, 1718

\bibitem[\protect\citeauthoryear{Miniutti et al.}{2019}]{Miniutti.2019} Miniutti G., Saxton R.~D., Giustini M., Alexander K.~D., Fender R.~P., Heywood I., Monageng I., et al., 2019, Natur, 573, 381

\bibitem[\protect\citeauthoryear{Mizumoto et al.}{2021}]{Mizumoto.2021} Mizumoto M., Nomura M., Done C., Ohsuga K., Odaka H., 2021, MNRAS, 503, 1442

\bibitem[\protect\citeauthoryear{Ni et al.}{2018}]{Ni.2018} Ni Q., Brandt W.~N., Luo B., Hall P.~B., Shen Y., Anderson S.~F., Plotkin R.~M., et al., 2018, MNRAS, 480, 5184

\bibitem[\protect\citeauthoryear{Ohsuga \& Mineshige}{2011}]{Ohsuga.2011} Ohsuga K., Mineshige S., 2011, ApJ, 736, 2

\bibitem[\protect\citeauthoryear{Osterbrock \& Pogge}{1985}]{Osterbrock.1985} Osterbrock D.~E., Pogge R.~W., 1985, ApJ, 297, 166

\bibitem[\protect\citeauthoryear{Polletta et al.}{2007}]{Polletta.2007} Polletta M., Tajer M., Maraschi L., Trinchieri G., Lonsdale C.~J., Chiappetti L., Andreon S., et al., 2007, ApJ, 663, 81

\bibitem[\protect\citeauthoryear{Pounds, Done \& Osborne}{1995}]{Pounds.1995} Pounds K.~A., Done C., Osborne J.~P., 1995, MNRAS, 277, L5

\bibitem[\protect\citeauthoryear{Rakshit et al.}{2017}]{Rakshit.2017} Rakshit S., Stalin C.~S., Chand H., Zhang X.-G., 2017, ApJS, 229, 39

\bibitem[\protect\citeauthoryear{Parker et al.}{2021}]{Parker.2021} Parker M.~L., Alston W.~N., H{\"a}rer L., Igo Z., Joyce A., Buisson D.~J.~K., Chainakun P., et al., 2021, MNRAS, 508, 1798

\bibitem[\protect\citeauthoryear{Pancoast, Brewer \& Treu}{2011}]{Pancoast.2011} Pancoast A., Brewer B.~J., Treu T., 2011, ApJ, 730, 139

\bibitem[\protect\citeauthoryear{Rankine et al.}{2020}]{Rankine.2020} Rankine A.~L., Hewett P.~C., Banerji M., Richards G.~T., 2020, MNRAS, 492, 4553

\bibitem[\protect\citeauthoryear{Pancoast et al.}{2014}]{Pancoast.2014} Pancoast A., Brewer B.~J., Treu T., Park D., Barth A.~J., Bentz M.~C., Woo J.-H., 2014, MNRAS, 445, 3073

\bibitem[\protect\citeauthoryear{Peterson et al.}{2004}]{Peterson.2004} Peterson B.~M., Ferrarese L., Gilbert K.~M., Kaspi S., Malkan M.~A., Maoz D., Merritt D., et al., 2004, ApJ, 613, 682

\bibitem[\protect\citeauthoryear{Peterson}{2014}]{Peterson.2014} Peterson B.~M., 2014, SSRv, 183, 253

\bibitem[\protect\citeauthoryear{Richards et al.}{2006}]{Richards.2006} Richards G.~T., Lacy M., Storrie-Lombardi L.~J., Hall P.~B., Gallagher S.~C., Hines D.~C., Fan X., et al., 2006, ApJS, 166, 470

\bibitem[\protect\citeauthoryear{Richards et al.}{2011}]{Richards.2011} Richards G.~T., Kruczek N.~E., Gallagher S.~C., Hall P.~B., Hewett P.~C., Leighly K.~M., Deo R.~P., et al., 2011, AJ, 141, 167

\bibitem[\protect\citeauthoryear{Plotkin et al.}{2010}]{Plotkin.2010} Plotkin R.~M., Anderson S.~F., Brandt W.~N., Diamond-Stanic A.~M., Fan X., MacLeod C.~L., Schneider D.~P., et al., 2010, ApJ, 721, 562

\bibitem[\protect\citeauthoryear{Ponti et al.}{2012}]{Ponti.2012} Ponti G., Papadakis I., Bianchi S., Guainazzi M., Matt G., Uttley P., Bonilla N.~F., 2012, A\&A, 542, A83

\bibitem[\protect\citeauthoryear{Poutanen et al.}{2007}]{Poutanen.2007} Poutanen J., Lipunova G., Fabrika S., Butkevich A.~G., Abolmasov P., 2007, MNRAS, 377, 1187

\bibitem[\protect\citeauthoryear{Pu et al.}{2020}]{Pu.2020} Pu X., Luo B., Brandt W.~N., Timlin J.~D., Liu H., Ni Q., Wu J., 2020, ApJ, 900, 141

\bibitem[\protect\citeauthoryear{Reeves et al.}{2000}]{Reeves.2000} Reeves J.~N., O'Brien P.~T., Vaughan S., Law-Green D., Ward M., Simpson C., Pounds K.~A., et al., 2000, MNRAS, 312, L17

\bibitem[\protect\citeauthoryear{Reeves et al.}{2020}]{Reeves.2020} Reeves J.~N., Braito V., Chartas G., Hamann F., Laha S., Nardini E., 2020, ApJ, 895, 37

\bibitem[\protect\citeauthoryear{Ross \& Fabian}{2005}]{Ross.2005} Ross R.~R., Fabian A.~C., 2005, MNRAS, 358, 211

\bibitem[\protect\citeauthoryear{Schlafly \& Finkbeiner}{2011}]{Schlafly.2011} Schlafly E.~F., Finkbeiner D.~P., 2011, ApJ, 737, 103

\bibitem[\protect\citeauthoryear{Schlegel, Finkbeiner \& Davis}{1998}]{Schlegel.1998} Schlegel D.~J., Finkbeiner D.~P., Davis M., 1998, ApJ, 500, 525

\bibitem[\protect\citeauthoryear{Shakura \& Sunyaev}{1973}]{Shakura.1973} Shakura N.~I., Sunyaev R.~A., 1973, A\&A, 500, 33

\bibitem[\protect\citeauthoryear{Silva, Maiolino \& Granato}{2004}]{Silva.2004} Silva L., Maiolino R., Granato G.~L., 2004, MNRAS, 355, 973

\bibitem[\protect\citeauthoryear{Simpson et al.}{1999}]{Simpson.1999} Simpson C., Ward M., O'Brien P., Reeves J., 1999, MNRAS, 303, L23

\bibitem[\protect\citeauthoryear{S{\k{a}}dowski et al.}{2011}]{Sadowski.2011} S{\k{a}}dowski A., Abramowicz M., Bursa M., Klu{\'z}niak W., Lasota J.-P., R{\'o}{\.z}a{\'n}ska A., 2011, A\&A, 527, A17

\bibitem[\protect\citeauthoryear{Shemmer \& Netzer}{2002}]{Shemmer.2002} Shemmer O., Netzer H., 2002, ApJL, 567, L19

\bibitem[\protect\citeauthoryear{Shen et al.}{2011}]{Shen.2011} Shen Y., Richards G.~T., Strauss M.~A., Hall P.~B., Schneider D.~P., Snedden S., Bizyaev D., et al., 2011, ApJS, 194, 45

\bibitem[\protect\citeauthoryear{Shen \& Ho}{2014}]{Shen.2014} Shen Y., Ho L.~C., 2014, Natur, 513, 210

\bibitem[\protect\citeauthoryear{Takeuchi, Ohsuga \& Mineshige}{2014}]{Takeuchi.2014} Takeuchi S., Ohsuga K., Mineshige S., 2014, PASJ, 66, 48

\bibitem[\protect\citeauthoryear{Tatum et al.}{2012}]{Tatum.2012} Tatum M.~M., Turner T.~J., Sim S.~A., Miller L., Reeves J.~N., Patrick A.~R., Long K.~S., 2012, ApJ, 752, 94

\bibitem[\protect\citeauthoryear{Turner et al.}{2007}]{Turner.2007} Turner T.~J., Miller L., Reeves J.~N., Kraemer S.~B., 2007, A\&A, 475, 121

\bibitem[\protect\citeauthoryear{Verner et al.}{1996}]{Verner.1996} Verner D.~A., Ferland G.~J., Korista K.~T., Yakovlev D.~G., 1996, ApJ, 465, 487

\bibitem[\protect\citeauthoryear{V{\'e}ron-Cetty, Joly \& V{\'e}ron}{2004}]{Veron.2004} V{\'e}ron-Cetty M.-P., Joly M., V{\'e}ron P., 2004, A\&A, 417, 515

\bibitem[\protect\citeauthoryear{Vestergaard \& Peterson}{2006}]{Vestergaard.2006} Vestergaard M., Peterson B.~M., 2006, ApJ, 641, 689

\bibitem[\protect\citeauthoryear{Vignali et al.}{2000}]{Vignali.2000} Vignali C., Comastri A., Nicastro F., Matt G., Fiore F., Palumbo G.~G.~C., 2000, A\&A, 362, 69

\bibitem[\protect\citeauthoryear{Voges et al.}{1999}]{Voges.1999} Voges W., Aschenbach B., Boller T., Br{\"a}uninger H., Briel U., Burkert W., Dennerl K., et al., 1999, A\&A, 349, 389

\bibitem[\protect\citeauthoryear{Wang \& Netzer}{2003}]{Wang.2003} Wang J.-M., Netzer H., 2003, A\&A, 398, 927

\bibitem[\protect\citeauthoryear{Watarai et al.}{2000}]{Watarai.2000} Watarai K.-. ya ., Fukue J., Takeuchi M., Mineshige S., 2000, PASJ, 52, 133

\bibitem[\protect\citeauthoryear{Wilkins et al.}{2014}]{Wilkins.2014} Wilkins D.~R., Kara E., Fabian A.~C., Gallo L.~C., 2014, MNRAS, 443, 2746

\bibitem[\protect\citeauthoryear{Willingale et al.}{2013}]{Willingale.2013} Willingale R., Starling R.~L.~C., Beardmore A.~P., Tanvir N.~R., O'Brien P.~T., 2013, MNRAS, 431, 394

\bibitem[\protect\citeauthoryear{Wilms, Allen \& McCray}{2000}]{Wilms.2000} Wilms J., Allen A., McCray R., 2000, ApJ, 542, 914

\bibitem[\protect\citeauthoryear{Wolfire et al.}{2003}]{Wolfire.2003} Wolfire M.~G., McKee C.~F., Hollenbach D., Tielens A.~G.~G.~M., 2003, ApJ, 587, 278

\bibitem[\protect\citeauthoryear{Woo et al.}{2015}]{Woo.2015} Woo J.-H., Yoon Y., Park S., Park D., Kim S.~C., 2015, ApJ, 801, 38

\bibitem[\protect\citeauthoryear{Wu et al.}{2011}]{Wu.2011} Wu J., Brandt W.~N., Hall P.~B., Gibson R.~R., Richards G.~T., Schneider D.~P., Shemmer O., et al., 2011, ApJ, 736, 28

\bibitem[\protect\citeauthoryear{Wu et al.}{2012}]{Wu.2012} Wu J., Brandt W.~N., Anderson S.~F., Diamond-Stanic A.~M., Hall P.~B., Plotkin R.~M., Schneider D.~P., et al., 2012, ApJ, 747, 10

\bibitem[\protect\citeauthoryear{Yang et al.}{2021}]{Yang.2021} Yang J., Paragi Z., Nardini E., Baan W.~A., Fan L., Mohan P., Varenius E., et al., 2021, MNRAS, 500, 2620

\bibitem[\protect\citeauthoryear{York et al.}{2000}]{York.2000} York D.~G., Adelman J., Anderson J.~E., Anderson S.~F., Annis J., Bahcall N.~A., Bakken J.~A., et al., 2000, AJ, 120, 1579

\bibitem[\protect\citeauthoryear{Zhou et al.}{2006}]{Zhou.2006} Zhou H., Wang T., Yuan W., Lu H., Dong X., Wang J., Lu Y., 2006, ApJS, 166, 128

\bibitem[\protect\citeauthoryear{Zhou et al.}{2010}]{Zhou.2010} Zhou X.-L., Zhang S.-N., Wang D.-X., Zhu L., 2010, ApJ, 710, 16

\bibitem[\protect\citeauthoryear{Zhu, Zhang \& Tang}{2009}]{Zhu.2009} Zhu L., Zhang S.~N., Tang S., 2009, ApJ, 700, 1173

\end{thebibliography}




\appendix

\section{Optical/UV Spectral Analysis}
\label{sec-optuvfit}

\subsection{Optical Spectral Analysis}
\label{sec-opt-spec}
The optical spectrum of \rxj0134\ resembles a typical NLS1 galaxy. In order to determine the parameters of various emission lines and continuum, we perform a detailed fitting to the optical spectrum. Firstly, the spectrum is corrected for the Galactic dust reddening with $E(B-V) = 0.0144$ (\citealt{Schlafly.2011}) and the extinction model in \citet{Fitzpatrick.2007}. Then the wavelength and flux are de-redshifted to AGN's rest frame with $z=0.237$. Then we adopt the prescription described in \citet{Dong.2008} to fit the spectrum.

The model of \citet{Dong.2008} comprises a power law with a slope of -2.5 for the underlying continuum, and an Fe {\sc ii} template from \citet{Veron.2004}. Since no stellar absorption features can be identified in the spectrum, star lights from host galaxy should not affect our emission line fitting, and so we do not include a host galaxy component. This is also confirmed by the extremely weak host galaxy component in the best-fit SED model in Figure~\ref{fig-sedfit}. All the narrow emission lines are fitted with Gaussian profiles of the same shape. [O {\sc iii}]$\lambda$5007 is fitted with two Gaussian profiles, one for the central component, and the other for the blue-shifted component. The intensity ratios of emission line doublets such as [O {\sc iii}]$\lambda$4959/5007 and [N {\sc ii}]$\lambda$6550/6585 are fixed at their theoretical values. For the broad Balmer lines, we first try Lorentzian profiles to fit the broad component (\citealt{Zhou.2006, Goad.2012, Rakshit.2017}). The velocity shifts and line widths of H$\alpha$ and H$\beta$ are kept the same, but their fluxes are left as free parameters. This model configuration assumes that different line components can have different Balmer decrements, which is possible because different emission line regions may have different electron densities and ionization states (e.g. \citealt{Kwan.1979, Mathews.1980, Jin.2012a, Jin.2012b}).

\begin{figure}
\centering
\includegraphics[trim=0.15in 0.4in 0.0in 0.1in, clip=1, scale=0.485]{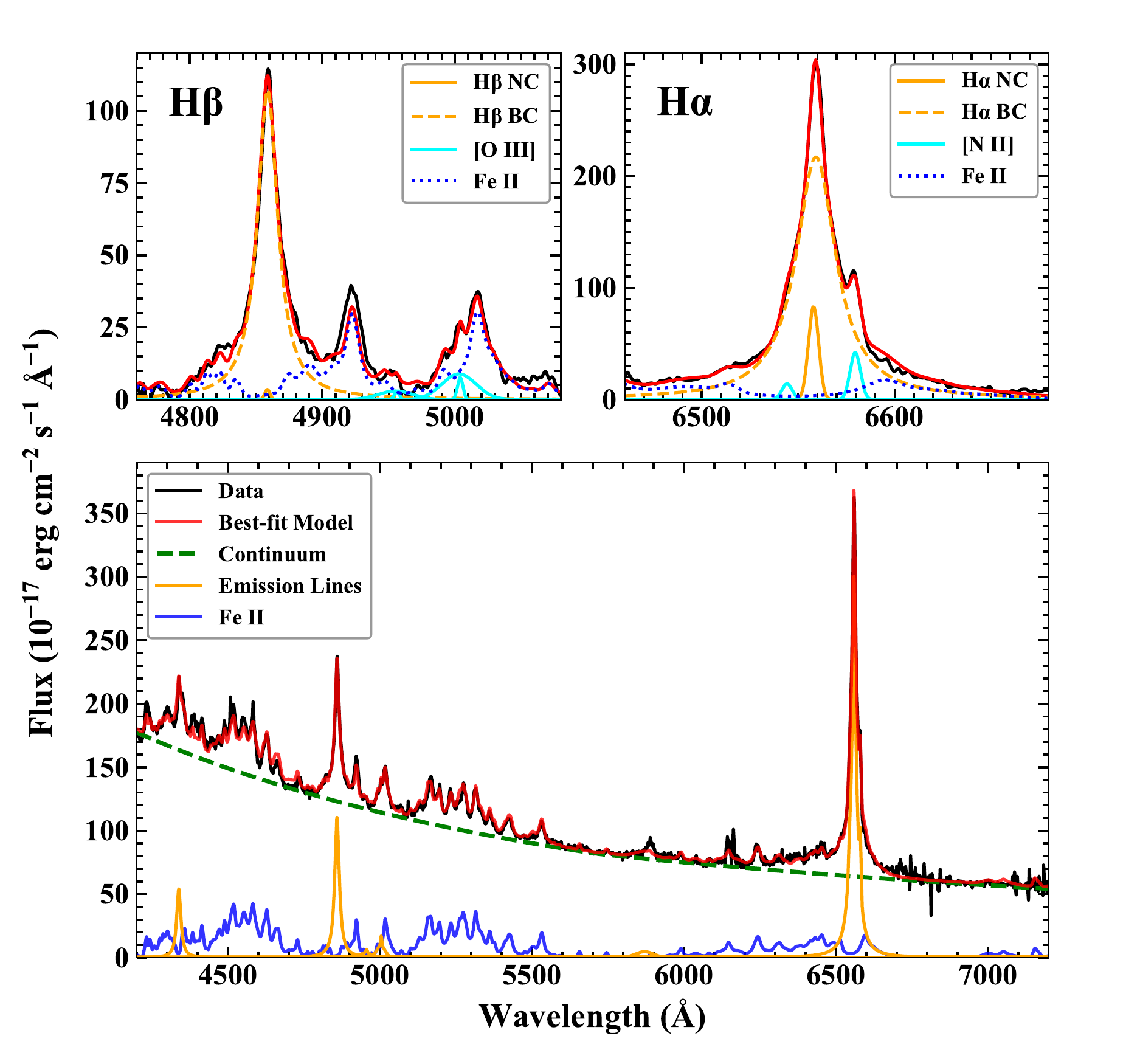} 
\caption{The de-reddened and de-redshifted optical spectrum (black) of \rxj0134\ observed by SSO in 2019-12-19, and the best-fit model (red) with multiple components including the intrinsic accretion disc continuum, Fe {\sc ii} emission, and multiple Gaussian/Lorentzian profiles to fit various strong emission lines. No significant host galaxy contribution is found in the spectrum.}
\label{fig-optfit}
\end{figure}

Figure~\ref{fig-optfit} shows the fitting results for the entire optical spectrum. All the line parameters are listed in Table~\ref{tab-optfit}. We can see that Balmer lines (e.g. H$\alpha$, H$\beta$) are well fitted by this Lorentzian decomposition. The FWHM of the broad component is 1140 $\pm$ 20 km s$^{-1}$, which is consistent with the value reported by \citet{Grupe.2010}. The rest-frame equivalent width (REW) of the narrow component (NC) of H$\beta$ is only 0.1 $\pm$ 0.5 \AA, and its Balmer decrement is 28$^{+18}_{-3}$. In comparison, the Balmer decrement of the broad component (BC) is 2.76 $\pm$ 0.16, which is a typical value of AGN's broad line region (BLR, e.g. \citealt{Shen.2011, Jin.2012b, Shen.2014}). A possible explanation is that at least part of the NC flux is from a region in the host galaxy where the dust extinction is severe, thus it contributes a lot more NC flux in H$\alpha$ than H$\beta$. One way to test this is to check other hydrogen lines such as Pa$\alpha$ in the near infrared, but no spectral data is available in this band.

\begin{figure}
\centering
\includegraphics[trim=0.15in 0.4in 0.0in 0.1in, clip=1, scale=0.465]{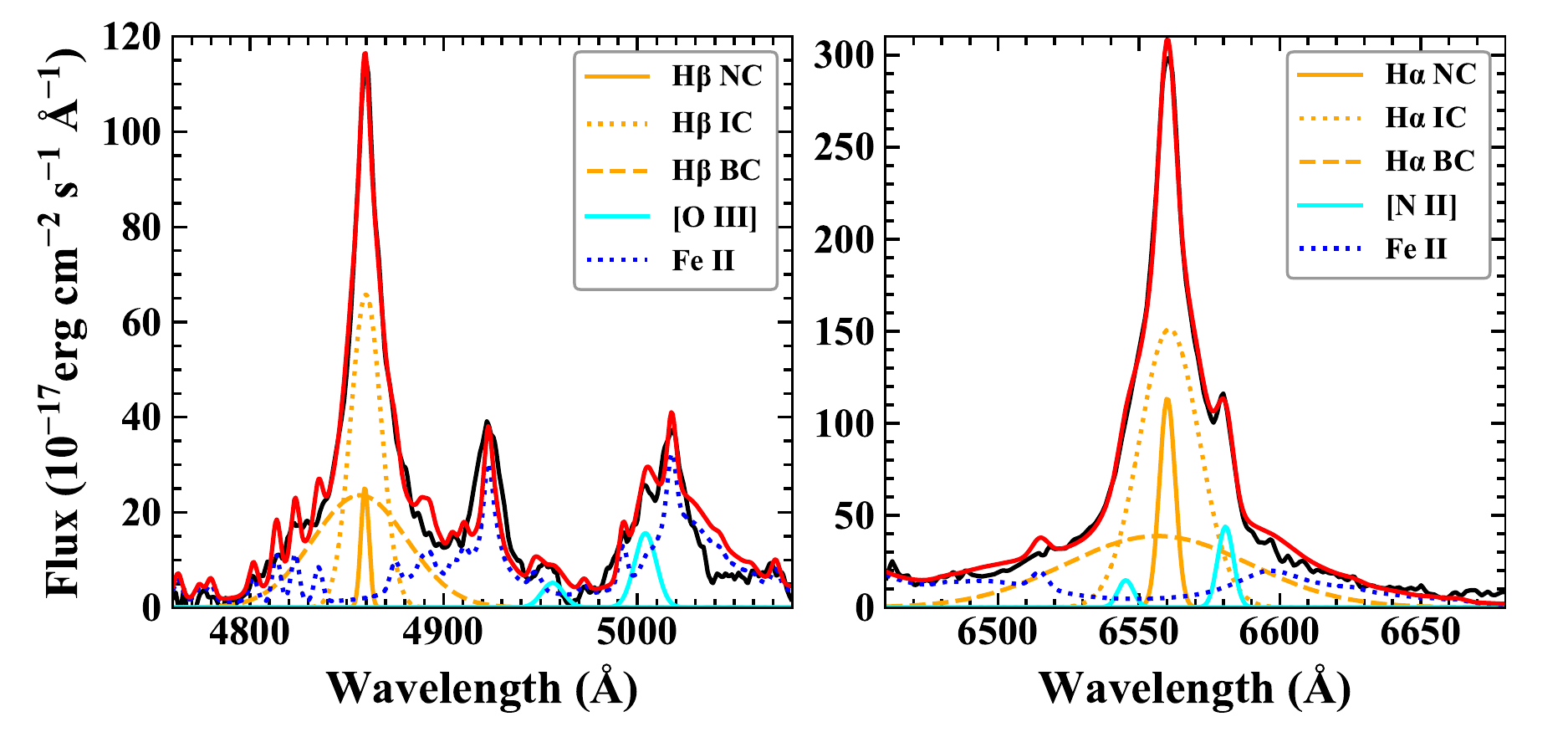} 
\caption{Fitting the H$\alpha$ and H$\beta$ lines with multiple Gaussian profiles.}
\label{fig-gaussfit}
\end{figure}

However, it is also possible that the large Balmer decrement of NC is a consequence of line decomposition, which is often ambiguous. As a further test, we follow the procedure of \citet{Liu.2019} to fit the Balmer lines by replacing the BC Lorentzian profile with two broad Gaussian profiles, i.e. an intermediate component (IC) and a BC (e.g. \citealt{Zhu.2009, Jin.2012a}). As shown in Figure~\ref{fig-gaussfit}, this model can also provide good fits to the Balmer lines. The FWHMs of NC, IC and BC are found to be 300 $\pm$ 20, 1140 $\pm$ 70 and 3520 $\pm$ 170 km s$^{-1}$, respectively. The Balmer decrement is 5.7 $\pm$ 1.8, 2.9 $\pm$ 0.4 and 2.1 $\pm$ 0.3 for the three components. Thus NC also appears heavily reddened. In this case, the ratio of H$\beta$ NC/[O{\sc iii}]$\lambda$5007 is 0.6 $\pm$ 0.4, which is much larger than 0.06 $\pm$ 0.11 as found in the Lorentzian fit and in most AGN (e.g. \citealt{Shen.2014}). Hence it is possible that the NC flux is overestimated in this line decomposition, and so we slightly prefer the Lorentzian method. The FWHM of H$\beta$ after subtracting the NC is 1410 $\pm$ 70 km s$^{-1}$, which is slightly broader than in the Lorentzian decomposition. Therefore, there is indeed some ambiguity in the width of Balmer lines caused by different line decomposition methods.

Our spectral fits also confirm the existence of strong Fe {\sc ii} emission, the REW of which is $41.5\pm3.4$ \AA\ within 4434 -- 4684 \AA. The flux ratio between the Fe {\sc ii} line in 4434 -- 4684 \AA\ and the broad H$\beta$ line is $R_{\rm FeII}=1.74\pm0.16$. After removing Fe{\sc ii} lines, [O{\sc iii}]$\lambda$5007 is found to be extremely weak, with the total REW being only 2.7 $\pm$ 0.8 \AA.

\subsection{UV Spectral Analysis}
\label{sec-uv-spec}
\begin{figure}
\centering
\includegraphics[trim=0.0in 0.3in 0.0in 0.1in, clip=1, scale=0.56]{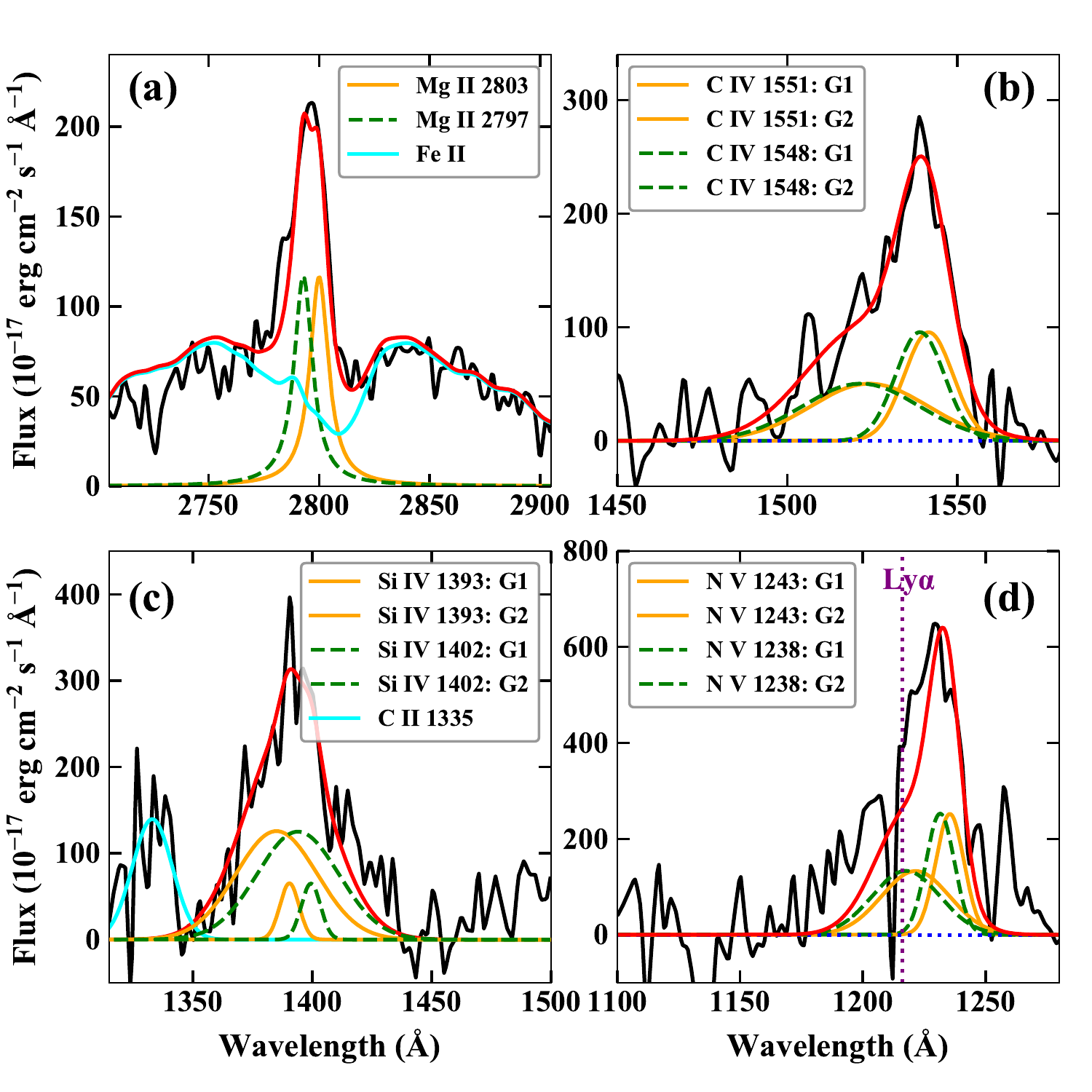} 
\caption{Fitting the Mg {\sc ii} $\lambda$2797/2803, C {\sc iv} $\lambda$1548/1551, N {\sc v} $\lambda$1238/1243 lines with multiple components. G1 and G2 indicate the two Gaussian components for the line's core region and blue wing. All the spectra are continuum-subtracted.}
\label{fig-uvfit}
\end{figure}

A special property of \rxj0134\ is its weak and strongly blue-shifted UV emission lines. We perform local spectral fitting to individual UV lines, such as the Mg {\sc ii} $\lambda$2797/2803, C {\sc iv} $\lambda$1548/1551, N {\sc v} $\lambda$1238/1243 doublets. The fits are plotted in Figure~\ref{fig-uvfit}, and the line parameters are listed in Table~\ref{tab-uvfit}.

The Mg {\sc ii} $\lambda$2797/2803 doublet can be well fitted with two Lorentzian components of the same shape and flux, as shown in Figure~\ref{fig-uvfit}a. The velocity shift is found to be -380 $\pm$ 90 km s$^{-1}$, indicating that the line is slightly blue-shifted. The FWHM is found to be 1170 $\pm$ 240 km s$^{-1}$, which is similar to the width of H$\beta$.

The C {\sc iv} line is much more blue-shifted and extended. It contains two adjacent lines at 1548 and 1551 \AA. Each of the two lines are fitted with two Gaussian components, including a narrow core component and a broad and blue-shifted wing component. As shown in Figure~\ref{fig-uvfit}b, this model can fit the line very well. The core component has a FWHM of 3250 $\pm$ 320 km s$^{-1}$, much broader than the BC of H$\beta$, and its velocity shift is -1780 $\pm$ 70 km s$^{-1}$. The wing component has a velocity shift of -5160 $\pm$ 660 km s$^{-1}$ and a much larger FWHM of 7800 $\pm$ 800 km s$^{-1}$. These two components have a total REW of 3.8 $\pm$ 1.5 \AA. The total REW of the C {\sc iv} doublet is much smaller than that measured from the major AGN population (e.g. 30 \AA\ in the quasars' composite spectrum: \citealt{Luo.2015, Coatman.2016}). The weak and blue-shifted C {\sc iv} doublet resembles WLQs.

The Si {\sc iv} $\lambda$1393/1402 doublet has a different shape from C {\sc iv}. The velocity shift of the core component is -710 $\pm$ 230 km s$^{-1}$, and its FWHM is 860 $\pm$ 320 km s$^{-1}$. The broad component is shifted by -1890 $\pm$ 20 km s$^{-1}$, with a FWHM of 3660 $\pm$ 40 km s$^{-1}$. Therefore, the line width and velocity shift of Si {\sc iv} are less extreme than C {\sc iv}. This is consistent with the fact that the ionization energy of Si {\sc iv} is lower, and so its emission region may have a larger radius, where the radiation pressure is weaker and the outflow speed is smaller.

The N {\sc v} $\lambda$1238/1243 doublet contains two broad lines. These two lines are often heavily blended with the broad Ly$\alpha$ line at 1216 \AA, making them difficult to decompose. We assume that N {\sc v} has the same line profile as C {\sc iv}, and then fit only the red side of the line hump within 1230 -- 1250 \AA. As shown in Figure~\ref{fig-uvfit}d, the result indicates that the line blend within 1150 -- 1250 \AA\ mostly come from the strong N {\sc v} $\lambda$1238/1243 doublet. The flux ratio of N {\sc v}/C {\sc iv} is 2.2 $\pm$ 0.4 , which is larger than the typical AGN value of less than unity (e.g. \citealt{Shemmer.2002}). This suggests that either the outflow in \rxj0134\ is more metal rich, or the flux of N {\sc v} in the line blend is over-estimated. Unfortunately, the spectral quality of this waveband is not good enough for a more accurate decomposition of Ly$\alpha$ and N {\sc v}.

\begin{table}
\centering
   \caption{Best-fit parameters of some optical emission lines shown in Figures~\ref{fig-optspec1}. {\it tied} means the value is tied to the corresponding component of H$\alpha$ during the spectral fitting. Errors represent 1$\sigma$ confidence limits. }
     \begin{tabular}{@{}lccc@{}}
     \hline
    Component & Parameter & Value & Unit \\
     \hline
    \multicolumn{4}{c}{H$\alpha$} \\
    Narrow Gauss & $v_{\rm line}$ & -260 $\pm$ 20 & km s$^{-1}$ \\
      & FWHM & 280 $\pm$ 50 & km s$^{-1}$ \\
      & Flux & 560 $\pm$ 160 & $10^{-17}$erg cm$^{-2}$ s$^{-1}$ \\
      & REW & 8.6 $\pm$ 2.5 &  \AA  \\
    Broad Lorentz & $v_{\rm line}$ & -250 $\pm$ 10 & km s$^{-1}$ \\
      & FWHM & 1140 $\pm$ 20 & km s$^{-1}$ \\
      & Flux & 8540 $\pm$ 240 & $10^{-17}$erg cm$^{-2}$ s$^{-1}$ \\
      & REW & 131.3 $\pm$ 3.6 &  \AA  \\
      
    \multicolumn{4}{c}{H$\beta$} \\
    Narrow Gauss & $v_{\rm line}$ & -260 {\it tied}  & km s$^{-1}$ \\
      & FWHM & 280 {\it tied}  & km s$^{-1}$ \\
      & Flux & 20 $\pm$ 70 & $10^{-17}$erg cm$^{-2}$ s$^{-1}$ \\
      & REW & 0.1 $\pm$ 0.5 &  \AA  \\
    Broad Lorentz & $v_{\rm line}$ & -250 {\it tied}  & km s$^{-1}$ \\
      & FWHM & 1140 {\it tied}  & km s$^{-1}$ \\
      & Flux & 3090 $\pm$ 160 & $10^{-17}$erg cm$^{-2}$ s$^{-1}$ \\
      & REW & 24.7 $\pm$ 1.2 &  \AA  \\
      
    \multicolumn{4}{c}{$[$O {\sc iii}$]~\lambda$5007} \\
    Core Gauss & $v_{\rm line}$ & -260 {\it tied} & km s$^{-1}$ \\
      & FWHM & 280 {\it tied} & km s$^{-1}$ \\
      & Flux & 40 $\pm$ 40 & $10^{-17}$erg cm$^{-2}$ s$^{-1}$ \\
      & REW & 0.3 $\pm$ 0.3 &  \AA  \\
    Wing Gauss & $v_{\rm line}$ & -310 $\pm$ 290 & km s$^{-1}$ \\
      & FWHM & 770 $\pm$ 270 & km s$^{-1}$ \\
      & Flux & 290 $\pm$ 90 & $10^{-17}$erg cm$^{-2}$ s$^{-1}$ \\
      & REW & 2.4 $\pm$ 0.7 &  \AA  \\
    \hline
     \end{tabular}
\label{tab-optfit}
\end{table}

\begin{table}
\centering
   \caption{Best-fit parameters of some UV emission lines shown in Figures~\ref{fig-optspec1}. For the Mg {\sc ii} $\lambda$2797/2803, C {\sc iv} $\lambda$1548/1551 and Si {\sc iv} $\lambda$1393/1402 doublets, a line ratio of 1:1 is adopted, and the reported flux and REW are only for one line. The N {\sc v} $\lambda$1238/1243 doublet parameters are not listed here because the line is assumed to have the same shape as C {\sc iv} $\lambda$1548/1551, except that its flux is higher by a factor of 2.2 $\pm$ 0.4. Errors are 1$\sigma$ confidence limits.}
     \begin{tabular}{@{}lccc@{}}
     \hline
    Component & Parameter & Value & Unit \\
     \hline      
    \multicolumn{4}{c}{Mg {\sc ii} $\lambda$2797} \\
     Single Lorentz & $v_{\rm line}$ & -380 $\pm$ 90 & km s$^{-1}$ \\
      & FWHM & 1170 $\pm$ 240 & km s$^{-1}$ \\
      & Flux & 2480 $\pm$ 610 & $10^{-17}$erg cm$^{-2}$ s$^{-1}$ \\
      & REW & 6.1 $\pm$ 1.5 &  \AA  \\
    
    \multicolumn{4}{c}{C {\sc iv} $\lambda$1548} \\
    Core Gauss & $v_{\rm line}$ & -1780 $\pm$ 70 & km s$^{-1}$ \\
      & FWHM & 3250 $\pm$ 320 & km s$^{-1}$ \\
      & Flux & 1800 $\pm$ 350 & $10^{-17}$erg cm$^{-2}$ s$^{-1}$ \\
      & REW & 1.7 $\pm$ 0.9 & \AA \\
    Wing Gauss & $v_{\rm line}$ & -5160 $\pm$ 660 & km s$^{-1}$ \\
      & FWHM & 7800 $\pm$ 800 & km s$^{-1}$ \\
      & Flux & 2280 $\pm$ 400 & $10^{-17}$erg cm$^{-2}$ s$^{-1}$ \\
      & REW & 2.1 $\pm$ 1.2 &  \AA  \\
      
    \multicolumn{4}{c}{Si {\sc iv} $\lambda$1393} \\
    Core Gauss & $v_{\rm line}$ & -710 $\pm$ 230 & km s$^{-1}$ \\
      & FWHM & 860 $\pm$ 320 & km s$^{-1}$ \\
      & Flux & 530 $\pm$ 300 & $10^{-17}$erg cm$^{-2}$ s$^{-1}$ \\
      & REW & 0.5 $\pm$ 0.2 & \AA \\
    Wing Gauss & $v_{\rm line}$ & -1890 $\pm$ 20 & km s$^{-1}$ \\
      & FWHM & 3660 $\pm$ 40 & km s$^{-1}$ \\
      & Flux & 4360 $\pm$ 530 & $10^{-17}$erg cm$^{-2}$ s$^{-1}$ \\
      & REW & 3.7 $\pm$ 1.2 &  \AA  \\
    \hline
     \end{tabular}
\label{tab-uvfit}
\end{table}



\bsp	
\label{lastpage}
\end{document}